\documentclass[floats,floatfix,showpacs,amssymb,prd,twocolumn,superscriptaddress,nofootinbib,nolongbibliography,reprint]{revtex4-2}

\usepackage{verbatim,mathtools,needspace,enumitem,etoolbox,physics,microtype,afterpage}
\usepackage{graphicx,epsf,epsfig}
\usepackage{bm}
\usepackage{color}
\usepackage[unicode, colorlinks=true, linkcolor=linkcolor, citecolor=linkcolor, filecolor=linkcolor,urlcolor=linkcolor, pdfusetitle]{hyperref}
\usepackage{textcomp}
\usepackage{amsfonts,amsmath,amssymb,mathrsfs,gensymb}
\usepackage{natbib}
\usepackage{prd_macros}
\usepackage{array}
\usepackage{multirow}
\usepackage{float}
\usepackage{rotating,array}
\usepackage[normalem]{ulem}
\usepackage{dcolumn}
\usepackage{braket}
\usepackage{appendix}
\usepackage{placeins}
\usepackage{caption}
\usepackage{subcaption}
\usepackage{comment}
\usepackage{float}

\usepackage[dvipsnames, usenames]{xcolor}

\definecolor{linkcolor}{rgb}{0.0,0.3,0.5}
\usepackage[all]{hypcap}
\usepackage[T1]{fontenc}
\usepackage[utf8]{inputenc}
\usepackage{tabularx}

\newcolumntype{P}[1]{>{\centering\arraybackslash}p{#1}}

\usepackage [english]{babel}
\usepackage [autostyle, english = american]{csquotes}
\MakeOuterQuote{"}

\newcommand{\dallas}{\affiliation{Department of Physics, The University of Texas at Dallas, Richardson, Texas 75080, USA}}

\newcommand\orcid[1]{\href{https://orcid.org/#1}{$\!$\includegraphics[scale=0.006]{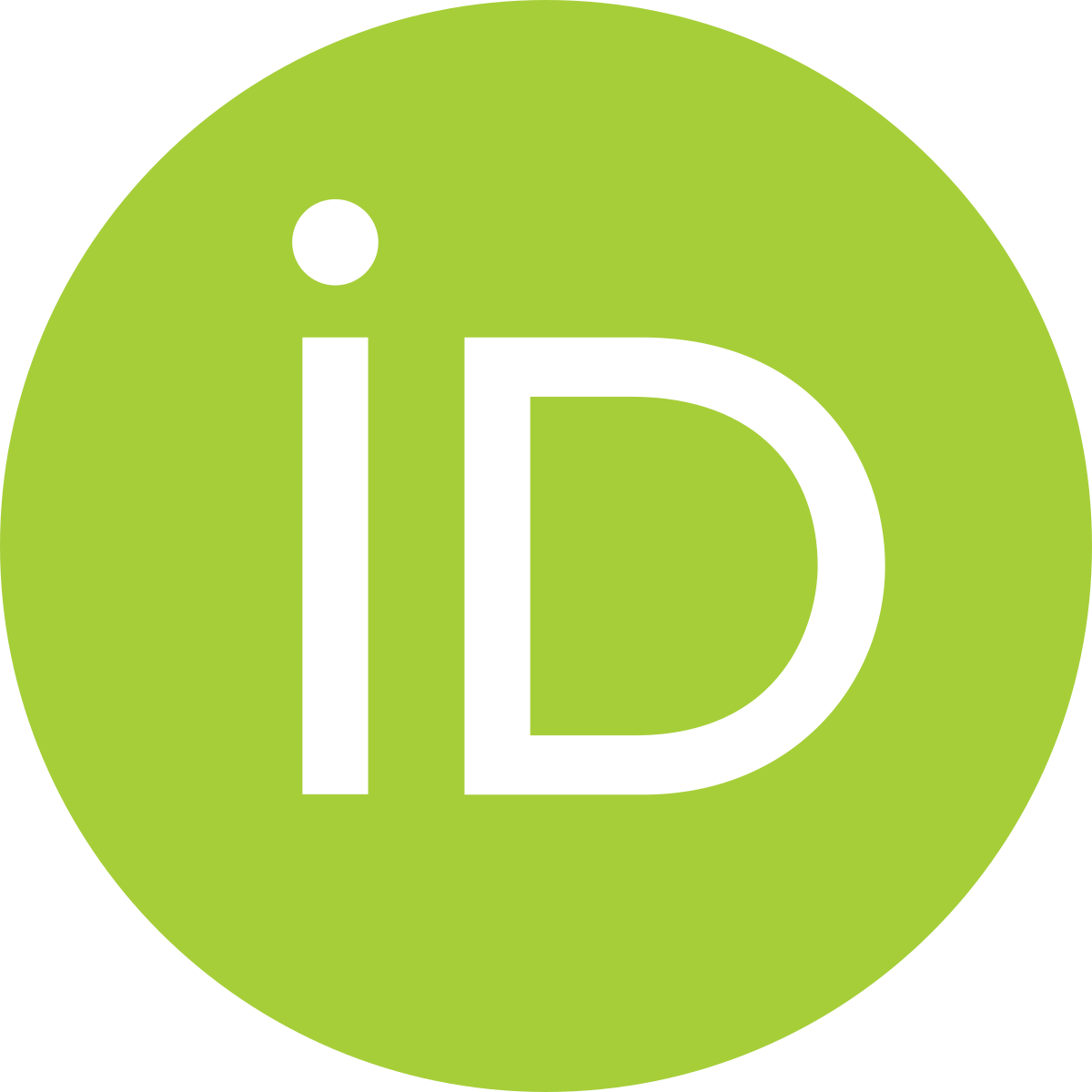} $\!\!$}}

\begin{document}



\title{Distribution of orbital inclinations for tidal disruption events by Kerr black holes}

\author{Tamanjyot Singh \orcid{0009-0006-4040-4407}}
\email{tamanjyot@utdallas.edu} 
\dallas

\author{Michael Kesden \orcid{0000-0002-5987-1471}}
\email{kesden@utdallas.edu}
\dallas

\date{\today}

\begin{abstract}

The Kerr metric that describes the spacetime of a spinning supermassive black hole (SMBH) is axisymmetric, implying that the nearly parabolic geodesics on which stars approach the SMBH depend on the inclination angle $\iota$ of the orbital angular momentum with respect to the SMBH spin.  This inclination affects both the geodesic deviation that determines whether a star is tidally disrupted and whether the tidal debris survives direct capture by the event horizon to produce an observable tidal disruption event (TDE).  The steady-state TDE rate is the rate at which stars are scattered into the loss cone determined by these spin- and inclination-dependent effects.  As the anisotropy of this loss-cone refilling is highly uncertain, we consider the two extreme limits in which stellar inclination is preserved (IP) or isotropized (ISO).  We calculate the inclination distribution in these two limits and find a prograde bias in the IP limit because of the strong retrograde bias for direct capture.  However, we find a retrograde bias in the ISO limit for intermediate SMBH masses when the empty loss cone suppresses capture and allows the weaker retrograde bias of geodesic deviation to dominate. Partially empty loss cones lead to steeper distributions of the penetration factor $\beta$ than for a full loss cone, with this effect even more pronounced in the ISO limit. We also calculate the total TDE rates and maximum SMBH mass $M_{\rm \bullet, max}$ for tidal disruption in these two limits.  In the IP limit, we find a highly spin-dependent capture cutoff in the TDE rate and $M_{\rm \bullet, max} \approx 10^{8.45} M_\odot$ for maximal SMBH spin.  In the ISO limit, we find a strong spin-dependent enhancement in the TDE rate at intermediate SMBH masses, a weakly spin-dependent capture cutoff above $M_\bullet \approx 10^{7.5} M_\odot$, and $M_{\rm \bullet, max} \approx 10^{7.95} M_\odot$ for maximal SMBH spin.

\end{abstract}

\maketitle

\section{Introduction} \label{S:Intro}

Stars that stray too close to a supermassive black hole (SMBH) may experience a tidal disruption event (TDE) in which they are partially or fully disrupted by the strong tidal forces exerted by the SMBH \cite{Wheeler1971,Hills1975}.  Emission by the resulting stellar debris as it is being accreted by the SMBH could power bright flares across the electromagnetic spectrum \cite{Rees1988,Phinney1989,Roth2020}.  Many TDE candidates have been found in X-ray and UV/optical surveys; for a recent review of these observations see \citeauthor{Gezari2021}~\cite{Gezari2021}. Several additional TDE candidates have been discovered after the publication of this review in X-ray \cite{Brightman2021, Sazonov2021}, UV/optical \cite{Hinkle2021, vanVelzen2021}, and infrared \cite{Wang2018, Jiang2021} surveys, and thousands more may soon be discovered in upcoming surveys \cite{Gezari2021}.

Predictions for TDE rates and the orbital properties of tidally disrupted stars are essential for the proper interpretation of these observed events.  Early theoretical work \cite{FrankRees1976,LightmanShapiro1977,CohnKulsrud1978} indicated that steady-state TDE rates are determined by the rate at which stellar scattering causes stars to diffuse onto low angular-momentum "loss-cone" orbits with pericenters within the tidal radius.  This theory was applied to local galaxy catalogs to predict TDE rates both for individual galaxies and per unit of cosmological volume \cite{MagorrianTremaine1999,WangMerritt2004,StoneMetzger2016}.  These predicted rates seem broadly consistent with the observed volumetric TDE rate \cite{vanVelzen2018}, although they fail to account for the observed enhancement of the TDE rate in post-starburst galaxies \cite{Arcavi2014,StonevanVelzen2016,Graur2018}.  See \citeauthor{StoneKesden2020}~\cite{StoneKesden2020} for a recent review on TDE rates and their astrophysical implications.

In addition to the total TDE rates, the distribution of orbital parameters for the tidally disrupted stars is also critical for predicting observational features of TDEs.  A necessary condition to produce an observable TDE is for the magnitude of the initial stellar orbital angular momentum $L$ to be large enough for the star to avoid direct capture by the SMBH event horizon ($L > L_{\rm cap}$), but small enough for tidal forces to strip appreciable mass from the stellar surface ($L < L_d$) \cite{Hills1975}.  Hydrodynamical simulations of the disruption process reveal that the peak accretion rate, time of peak accretion, amount of mass lost by the disrupted star, and asymptotic decay power-law index, inferred from the distribution in binding energy of the tidal debris, all depend on the orbital angular momentum $L$ \cite{Guillochon2013}.  Relativistic apsidal precession, which facilitates the stream-stream collisions that help transform tidal stream into accretion disks \cite{Rees1988,Kochanek1994,HayasakiStone2013,Shiokawa2015,Dai2015,Bonnerot2017,RossiKesden2021}, also depends on the orbital angular momentum, with the precession angle $\Delta\omega \propto L^{-2}$ at lowest post-Newtonian (PN) order.

SMBHs grow through gas accretion and binary mergers, both of which contribute angular momentum in addition to mass.  This implies that the SMBHs should have a non-zero dimensionless spin $\chi$ and be described in general relativity by the axisymmetric Kerr metric \cite{Kerr1963}.  In this case, the geodesics that describe stellar orbits will depend on both the spin $\chi$ and the inclination $\iota$, the asymptotic angle between the SMBH spin and the Newtonian orbital angular momentum vector, as will the angular-momentum thresholds for capture $L_{\rm cap}$ \cite{BardeenTeukolsky1972} and disruption $L_d$ \cite{ManasseMisner1963,Marck1983}.  In the absence of relativistic effects, the distribution of the angular momenta of tidally disrupted stars is expected to be isotropic (flat in $\cos\iota$).  This isotropy occurs because the semi-major axes of tidally disrupted stars are of the order of the influence radius $r_h \equiv GM_\bullet/\sigma^2$ for a SMBH of mass $M_\bullet$ in a galaxy with velocity dispersion $\sigma$, which is much greater than the gravitational radius $r_g \equiv GM_\bullet/c^2$ at which relativistic effects become dominant.

The distribution of TDE inclinations has several observational consequences.  For non-zero spin and inclination, nodal or Lense-Thirring precession \cite{LenseThirring1918} causes the line of ascending node of the tidal stream to precess, potentially delaying or preventing the stream from colliding with itself and circularizing \cite{Kochanek1994,StoneLoeb2012,Guillochon2015,BatraLu2021}.  If the stream evolves into a thin accretion disk that remains misaligned with the SMBH spin, the inclination determines the specific energy at the innermost stable circular orbit (ISCO) and thus the bolometric radiative efficiency.  For a maximally spinning SMBH ($\chi = 1$), this efficiency can vary with inclination by over an order of magnitude, from 42\% for a prograde ($\cos\iota = 1$) orbit to only 3.8\% on a retrograde ($\cos\iota = -1$) orbit \cite{BardeenTeukolsky1972}.  If the Bardeen-Petterson effect \cite{BardeenPetterson1975} drives the accretion disk into the equatorial plane of the SMBH, the inclination distribution will determine the fraction of TDEs with prograde/retrograde accretion disks.  Several TDEs have been observed to launch powerful jets \cite{Burrows2011,Cenko2012,Brown2015}, and their ability to do so through the Blandford-Znajek mechanism \cite{BlandfordZnajek1977} may depend on both the SMBH spin and TDE inclination \cite{2023arXiv230805161T}.  Finally, both the amplitude and frequency of quasi-periodic oscillations (QPOs) observed in TDE candidates \cite{Reis2012,Lin2015,Pasham2019} could depend on the SMBH spin and inclination \cite{StoneLoeb2012,vanVelzenPasham2021}.

To our knowledge, this is the first work explicitly focused on the distribution of TDE inclinations.  Early work implicitly considered the inclination distribution because of its influence on SMBH spin evolution, but this work either neglected the spin dependence of the tidal forces \cite{Young1977} or used a simplified treatment of loss-cone physics \cite{Beloborodov1992}.  More recently, \citeauthor{IvanovChernyakova2006}~\cite{IvanovChernyakova2006} calculated the inclination-dependent TDE cross sections, but neglected the loss-cone physics required to convert these cross sections into predicted TDE rates.  \citeauthor{Kesden2012}~\cite{Kesden2012} calculated total TDE rates using both spin-dependent tidal force and thresholds for direct capture, but the Monte Carlo simulations in this paper also assumed a full loss cone. \citeauthor{CoughlinNixon2022}~\cite{CoughlinNixon2022} calculated the distribution function of Boyer-Lindquist \cite{BoyerLindquist1967} radial coordinate at pericenter $r_p$ as a function of the SMBH mass and spin, again assuming a full loss cone.

\citeauthor{ServinKesden2017}~\cite{ServinKesden2017} self-consistently used relativistic tidal forces and direct-capture thresholds, and the loss-cone distribution function determined by them, to calculate both the total TDE rate and the distribution of angular momenta $L$ for Schwarzschild (non-spinning) SMBHs.  This paper extends that work to Kerr SMBHs, allowing for prediction of the TDE inclination distribution.

We briefly review how the thresholds for tidal disruption $L_d$ and direct capture $L_{\rm cap}$ are determined for the Kerr metric in Sec.~\ref{S:Disruption and Capture}.  We then examine how the spin and inclination dependence of these thresholds induces relativistic corrections to the differential TDE rates in Sec.~\ref{S:Loss Cone}.  We use these corrections to calculate TDE inclination distributions and total TDE rates in Secs.~\ref{S:Distribution of Inclinations} and \ref{S:Total Rates}.  We provide a brief synopsis of our results and discuss their implications in Sec.~\ref{S:Discussion}. 

\section{Disruption and capture} \label{S:Disruption and Capture} 

We assume that stars of mass $M_\star$ and radius $R_\star$ approach the SMBH on parabolic geodesics of the Kerr metric (specific energy $E = 1$ in units where the speed of light $c$ is unity).  This assumption is highly accurate because the Newtonian specific binding energy $\epsilon_\ast \approx \sigma^2$ of these stars is much less than the specific rest-mass energy $c^2$.  The stationarity and axisymmetry of the Kerr metric imply that parabolic geodesics can be characterized by three parameters: the component of the specific orbital angular momentum parallel to the symmetry axis $L_z$, the Carter constant $Q$ \cite{Carter1968}, and the Boyer-Lindquist \cite{BoyerLindquist1967} polar coordinate at pericenter $\theta_p$.  In the Newtonian limit far from the SMBH, the Carter constant reduces to the square of the magnitude of the component of the specific orbital angular momentum perpendicular to the symmetry axis.  This allows us to covariantly define the magnitude of the specific orbital angular momentum $L \equiv \sqrt{Q + L_z^2}$ and the inclination $\cos\iota \equiv L_z/L$.  In the Newtonian limit, $\cos\theta_p = \sin\iota\sin\omega$, where $\omega$ is the argument of pericenter.  The maximum tidal acceleration depends on $\theta_p$ at the sub-percent level \cite{Servin2018}, while direct capture is independent of $\theta_p$.  We neglect this dependence and calculate the angular-momentum thresholds for tidal disruption $L_d$ and direct capture $L_{\rm cap}$, both of which depend on the dimensionless SMBH spin $\chi$ and inclination $\iota$.

\subsection{Tidal Disruption} \label{SS:TD}

In general relativity, a star is tidally disrupted when the tidal acceleration given by the geodesic-deviation equation exceeds the star's self-gravity.  In Fermi normal coordinates, the tidal acceleration is given by
\begin{equation}
\frac{d^2X^{(i)}}{d\tau^2} = -{C^{(i)}}_{(j)} X^{(j)}~,
\end{equation}
where $\tau$ is the proper time along the central timelike geodesic on which the star's center of mass travels, $X^{(i)}$ are Cartesian spatial coordinates in a spatial hypersurface orthogonal to this central geodesic, and
\begin{equation} 
{C^{(i)}}_{(j)} \equiv {R^\beta}_{\mu\alpha\nu} {\lambda_\beta}^{(i)} {\lambda^\mu}_{(0)} {\lambda^\alpha}_{(j)} {\lambda^\nu}_{(0)}
\end{equation}
is the tidal tensor \cite{ManasseMisner1963,Marck1983}.  Greek (Latin) indices run over the four (three) spacetime (spatial) coordinates, ${R^\beta}_{\mu\alpha\nu}$ is the Riemann tensor, and $\{ {\lambda^\mu}_{(0)}, {\lambda^\mu}_{(i)} \}$ are the orthonormal tetrad of 4-vectors with respect to which the Fermi normal coordinates $\{ \tau, X^{(i)} \}$ are defined.  The tidal tensor is a real symmetric $3 \times 3$ matrix with three real eigenvalues and three corresponding orthogonal spatial eigenvectors.  One of these eigenvalues,
\begin{equation} \label{E:NegEig}
\xi_- = -\frac{2M_\bullet}{r^3} \left\{ 1 + \frac{3[L^2 - 2\chi M_\bullet L\cos\iota + (\chi M_\bullet)^2]}{2r^2} \right\}
\end{equation}
is negative leading to tidal stretching along the direction of the corresponding eigenvector.  The main-sequence stars we consider in this paper are well described by Newtonian gravity, so the gravitational acceleration at their surfaces is given by $\alpha_\star \equiv M_\star/{R_\star}^2$.  Tidal disruption occurs on geodesics for which the tidal acceleration at pericenter exceeds the star's self gravity,
\begin{equation} \label{E:DisThresh}
-\xi_- R_\star > 2\beta_d^3 \alpha_\star~,
\end{equation}
where $\beta_d$ is a factor of order unity that accounts for stellar structure.  For solar-type stars well described by a polytropic index $\gamma = 4/3$, $\beta_d \approx 1.9$ \cite{Guillochon2013}.  Eq.~(\ref{E:DisThresh}) determines the angular-momentum threshold $L_d(\chi, \iota)$ for tidal disruption which depends on both the SMBH spin $\chi$ and orbital inclination $\iota$ through Eq.~(\ref{E:NegEig}).

\begin{figure*}[t!]
\begin{subfigure}[b]{0.49\textwidth}
\includegraphics[width=\linewidth]{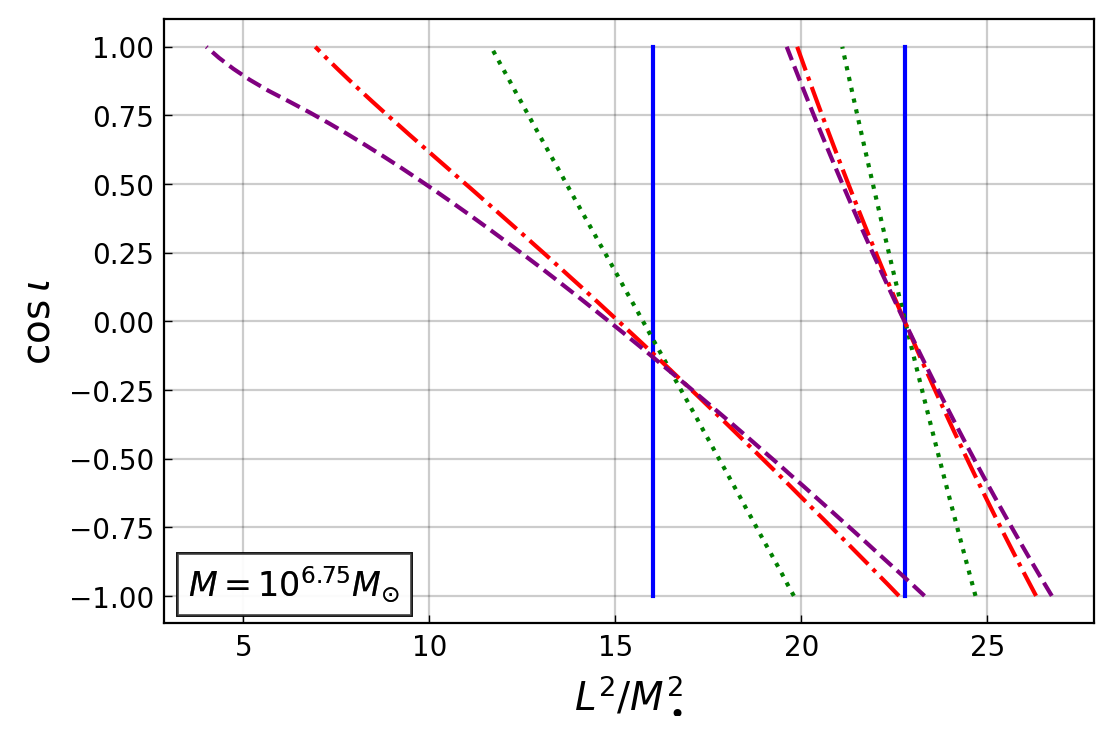}
\end{subfigure}
\begin{subfigure}[b]{0.49\textwidth}
\includegraphics[width=\linewidth]{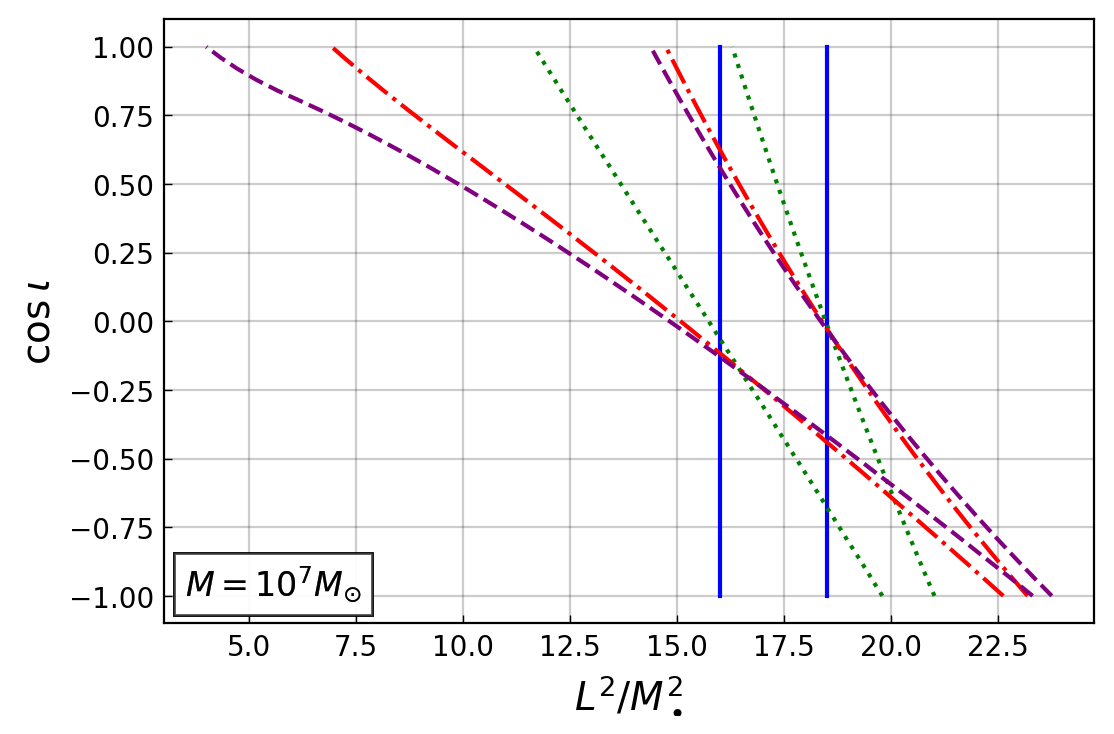}
\end{subfigure}
\begin{subfigure}[b]{0.49\textwidth}
\includegraphics[width=\linewidth]{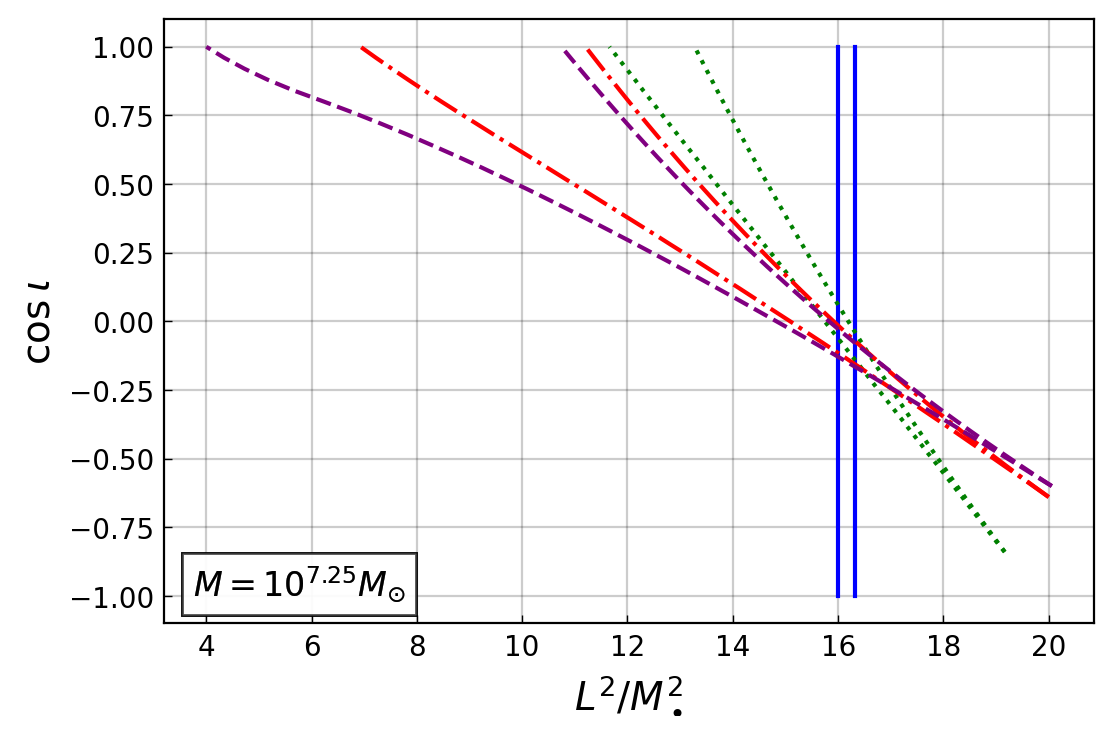}
\end{subfigure}
\begin{subfigure}[b]{0.49\textwidth}
\includegraphics[width=\linewidth]{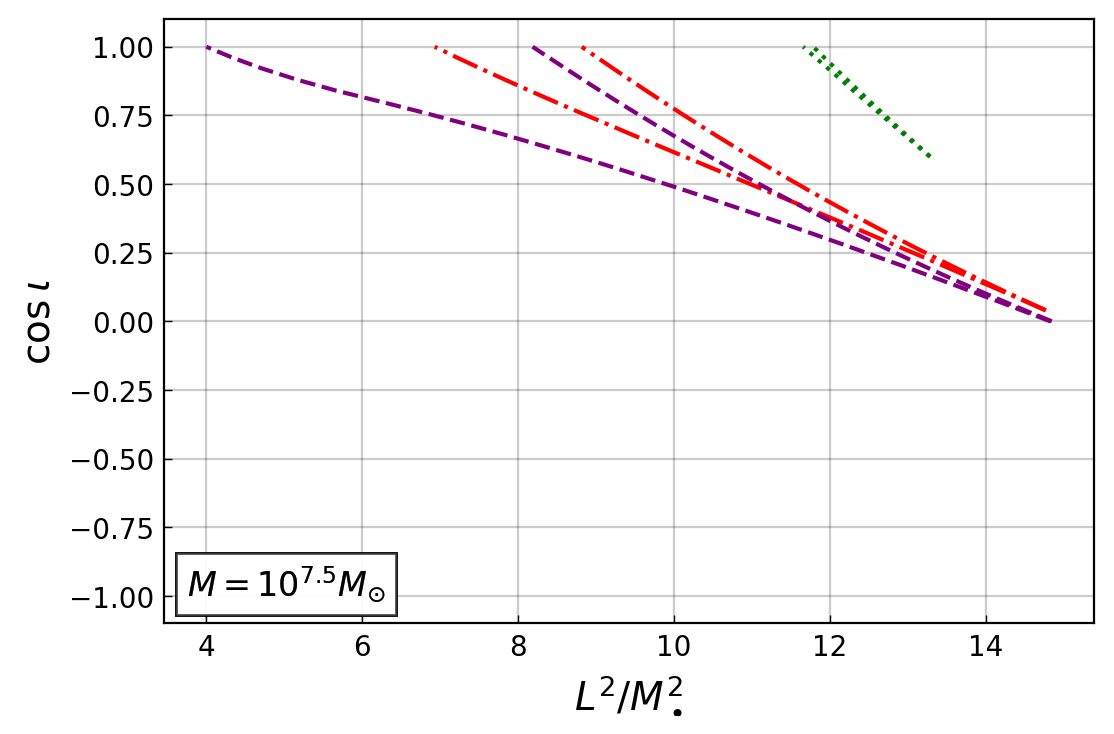}
\end{subfigure}
\caption{Angular-momentum thresholds for tidal disruption $L_d$ and direct capture $L_{\rm cap}$ as function of orbital inclination $\iota$.  The top left, top right, bottom left, and bottom right panels correspond to SMBH masses $M_\bullet = 10^{6.75}, 10^7, 10^{7.25},$ and $10^{7.5}\,M_\odot$ respectively.  Within each panel, the solid blue, dotted green, dot-dashed red, and dashed purple curves correspond to SMBH spins $\chi = 0, 0.5, 0.9,$ and 1, and the left (right) curve indicates $L_d$ ($L_{\rm cap}$).  The curves are not shown in the bottom right panel for spins and inclinations for which $L_d < L_{\rm cap}$, as TDEs are not possible in this portion of parameter space.
}
\label{fig: C&D Angular momenta}
\end{figure*}

\subsection{Direct Capture by the Event Horizon} \label{SS:cap}

The radial velocity along a parabolic ($E = 1$) Kerr geodesic is given by \cite{BardeenTeukolsky1972}
\begin{equation}
\Sigma^2 \left( \frac{dr}{d\tau} \right)^2 = V_r~,
\end{equation}
where $\Sigma \equiv r^2 + (\chi M_\bullet)^2 \cos^2\theta$ and
\begin{align} \label{E:RadPot}
V_r &= 2M_\bullet r[r^2 + L^2 - 2\chi M_\bullet L\cos\iota + (\chi M_\bullet)^2] \notag \\
& \quad - L^2[r^2 + (\chi M_\bullet)^2(1 - \cos^2\iota)]~.
\end{align}
The angular-momentum threshold $L_{\rm cap}(\chi, \iota)$ for direct capture by the event horizon is found by simultaneously solving the equations $V_r = 0$ and $dV_r/dr = 0$ simultaneously for $r$ and $L$.  This determines the upper bound $L_{\rm cap}(\chi, \iota)$ for direct captures in which the stellar debris is swallowed by the event horizon before it can emit appreciable electromagnetic radiation.

In Fig.~\ref{fig: C&D Angular momenta}, we show the two thresholds $L_d(\chi, \iota)$ and $L_{\rm cap}(\chi, \iota)$ between which observable TDEs are possible as functions of $\cos\iota$ for various SMBH masses $M_\bullet$ and spins $\chi$.  With the scaling $(L/M_\bullet)^2$ used in this figure, the capture curves $L_{\rm cap}$ are independent of SMBH mass and the TDE rate is proportional to the area between the $L_d$ and $L_{\rm cap}$ curves in the full loss-cone limit \cite{MagorrianTremaine1999}.  To lowest PN order, $(L_d/M_\bullet)^2 \propto {M_\bullet}^{-2/3}$, so the tidal disruption curves move to the left as SMBH mass increases from the top left to bottom right panels.  This reduces the area between the curves at large masses, resulting in a suppression of the TDE rate due to direct capture by the event horizon.  For nonzero spin, both tidal disruption and direct capture are biased towards retrograde ($\cos\iota < 0$) orbits, i.e. $L_d$ and $L_{\rm cap}$ are monotonically increasing functions of inclination $\iota$.  However, this bias is stronger for capture than disruption, leading to greater distance between the disruption and capture curves for prograde ($\cos\iota > 0$) orbits and a  prograde bias in observable TDEs in the full loss-cone limit.  In the next section, we will describe our treatment of loss-cone theory which determines whether this prograde bias persists for the partially empty loss cones relevant to the more massive SMBHs for which relativistic effects are significant.

\section{Loss-Cone Theory} \label{S:Loss Cone}


We begin by reviewing how TDE rates are calculated for loss cones in Newtonian gravity, then modify this calculation to account for relativistic effects.

\subsection{Newtonian Loss Cone}

To predict the distribution of TDE inclinations, we will modify the approach of \citeauthor{CohnKulsrud1978}~\cite{CohnKulsrud1978} as applied by \citeauthor{WangMerritt2004}~\cite{WangMerritt2004} to galactic centers described by singular isothermal spheres ($\rho = \sigma^2/2\pi Gr^2$).  We briefly describe this approach below; further details are provided in Secs.~6.1.2 and 6.1.3 of \citeauthor{MerrittBook}~\cite{MerrittBook}.

Eddington's formula implies that the isotropic distribution function for the singular isothermal sphere is
\begin{equation} \label{E:f_FLC}
f_{\rm FLC}(\mathcal{E}) = \left( \frac{M_\bullet}{M_\star} \right) \frac{g(\mathcal{E}^\ast)}{(\sigma r_h)^3}~,
\end{equation}
where $\mathcal{E}^\star \equiv \mathcal{E}/\sigma^2$ is the dimensionless specific binding energy and $g(\mathcal{E}^\ast) \approx \mathcal{E}^{\ast 1/2}/\sqrt{2}\pi^3$ for the regime $\mathcal{E}^\star \gtrsim 1$ that dominates the TDE rate.  In the steady-state solution of \citeauthor{WangMerritt2004}~\cite{WangMerritt2004}, the loss cone generated by the SMBH reduces the distribution function by a factor
\begin{equation} \label{E:f_elc}
\frac{f_{\rm ELC}(\mathcal{E}, \mathcal{R})}{f_{\rm FLC}(\mathcal{E})} = \frac{\ln(\mathcal{R}/\mathcal{R}_0(\mathcal{E}))}{\ln(1/\mathcal{R}_0(\mathcal{E}))}~,
\end{equation}
where $\mathcal{R} \equiv [L/L_{\rm c}(\mathcal{E}^\ast)]^2$ is normalized by $L_{\rm c}(\mathcal{E}^\ast)$, the angular momentum of a circular orbit of dimensionless specific binding energy $\mathcal{E}^\ast$, and $\mathcal{R}_0$ is the value of this variable below which the loss cone is completely empty.  This factor and the numerical value of $\mathcal{R}_0$ are found by solving the Fokker-Planck equation subject to the appropriate boundary conditions set by the loss cone; a good approximation is
\begin{equation} \label{E:R_0}
\mathcal{R}_0(\mathcal{E}) \approx \mathcal{R}_{\rm lc} e^{-(q^2 + q^4)^{1/4}}~,
\end{equation}
where $\mathcal{R}_{\rm lc}$ is the value of $\mathcal{R}$ at the boundary of the loss cone and
\begin{equation} \label{E:q}
q(\mathcal{E}^\ast) \approx \frac{20}{9} \ln\Lambda \left( \frac{M_\star}{M_\bullet} \right) \left( \frac{r_h}{r_{\rm lc}} \right) (\mathcal{E}^\ast)^{-2}
\end{equation}
is the ratio of the period of an orbit of energy $\mathcal{E}^\ast$ to the time it takes for a star to diffuse across the loss cone, i.e. for 
$\mathcal{R}$ to change by an amount $\mathcal{R}_{\rm lc}$.  The Coulomb logarithm is approximated by $\ln\Lambda \approx \ln (0.4M_\bullet/M_\star)$ and $r_{\rm lc} = r_t/\beta_d$ with $r_t \equiv (M_\bullet/M_\star)^{1/3}R_\star$.

The flux of stars into the loss cone per unit dimensionless specific binding energy and squared angular momentum is
\begin{align}
\frac{d^2\dot{N}}{d\mathcal{E}^\ast d(L^2/M_\bullet^2)} &= \left( \frac{2\pi\sigma GM_\bullet}{c} \right)^2 f(\mathcal{E}, \mathcal{R}) \label{E:doubdif} \\
&\approx \frac{2\sqrt{2}\sigma^5\mathcal{E}^{\ast 1/2}}{\pi GM_\star c^2} \label{E:doubdif_FLC} \\
&\approx \frac{2\sqrt{2}\sigma^5\mathcal{E}^{\ast 1/2}}{\pi GM_\star c^2} \frac{\ln(\mathcal{R}/\mathcal{R}_0)}{\ln(1/\mathcal{R}_0)}~, \label{E:doubdif_ELC}
\end{align}
where Eqs.~(\ref{E:doubdif_FLC}) and (\ref{E:doubdif_ELC}) apply to the full and empty loss cones respectively.  Integrating these equations with respect to the squared angular momentum yields the TDE rate per unit dimensionless specific binding energy:
\begin{align}
\frac{d\dot{N}}{d\mathcal{E}^\ast} &= \left( \frac{2\pi\sigma GM_\bullet}{c} \right)^2 \int f(\mathcal{E}, \mathcal{R})\, d(L^2/M_\bullet^2) \\
&\approx \frac{2\sqrt{2}\sigma^5\mathcal{E}^{\ast 1/2}}{\pi GM_\star c^2} A_{\rm FN} \label{E:dif_FLC} \\
&\approx \frac{2\sqrt{2}\sigma^5\mathcal{E}^{\ast 1/2}}{\pi GM_\star c^2} A_{\rm EN}(\mathcal{E}^\ast)~, \label{E:dif_ELC}
\end{align}
where the differential TDE rates for full and empty loss cones in the Newtonian limit are proportional to the factors
\begin{align}
A_{\rm FN} &\equiv \int_{0}^{L_{\rm lc}^2/M_\bullet^2} d(L^2/M_\bullet^2) = \left( \frac{L_{\rm lc}}{M_\bullet} \right)^2 \label{E:A_FN} \\ 
A_{\rm EN}(\mathcal{E}^\ast) &\equiv \int_{L_{0}^2/M_\bullet^2}^{L_{\rm lc}^2/M_\bullet^2} \frac{\ln(\mathcal{R}/\mathcal{R}_0)}{\ln(1/\mathcal{R}_0)}\, d(L^2/M_\bullet^2)~. \label{E:A_EN}
\end{align}

\begin{figure}[t!]
\includegraphics[width=\linewidth]{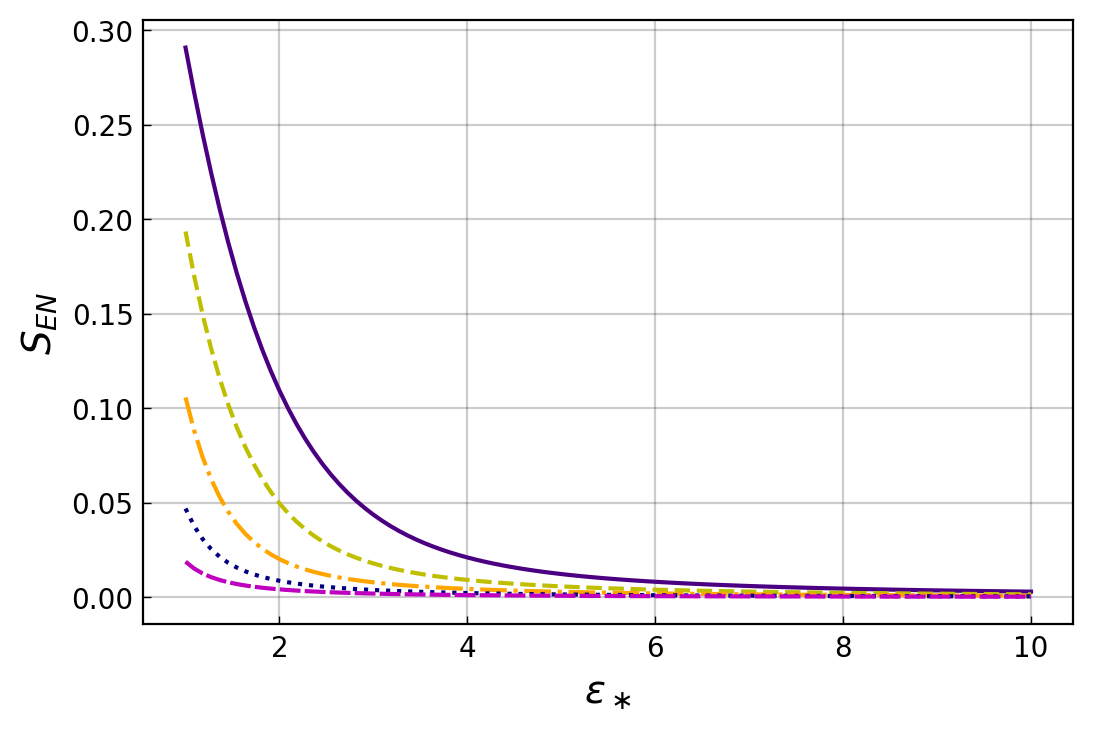}
\caption{The factor $S_{\rm EN}$ by which the TDE rate per unit dimensionless specific energy is suppressed by the emptiness of the loss cone in Newtonian gravity as a function of the dimensionless specific binding energy $\mathcal{E}^\ast$.  The solid purple, dashed yellow, dot-dashed orange, dotted blue, and dashed magenta curves correspond to SMBH masses of $M_\bullet = 10^6, 10^{6.5}, 10^7, 10^{7.5},$ and $10^8\, M_\odot$.
}
\label{fig:S_EN vs Epsilon}
\end{figure}

In Fig.~\ref{fig:S_EN vs Epsilon}, we plot the ratio $S_{\rm EN} \equiv A_{\rm EN}/A_{\rm FN}$ as a function of dimensionless specific binding energy $\mathcal{E}^\ast$.  We use the Newtonian value 
\begin{equation} \label{E:L_lc N}
    \left( \frac{L_{\rm lc}}{M_\bullet} \right)^2 = \frac{2r_tc^2}{GM_\bullet\beta_d} = 94.2\beta^{-1}_d \left( \frac{M_\bullet}{10^6 M_\odot} \right)^{-2/3}
\end{equation}

for the squared angular momentum at the boundary of the loss cone set by tidal disruption.  We adopt the $M_\bullet-\sigma$ relation
\begin{equation}
M_\bullet = M_{200} \left( \frac{\sigma}{200~{\rm km/s}} \right)^p~,
\end{equation}
between SMBH mass and host-galaxy velocity dispersion with the recent calibration $M_{200} = 10^{8.32}M_\odot,\, p = 5.64$ \cite{McConnellMa2013}.  The ratio $S_{\rm EN}$ monotonically decreases with $\mathcal{E}^\ast$ because more tightly bound orbits are less refilled by stellar diffusion over their shorter orbital periods.  This ratio also monotonically decreases with SMBH mass $M_\bullet$ because stellar densities and thus stellar diffusion are lower at the boundaries of larger loss cones.

\begin{figure*}
    \centering
    \includegraphics[width=\textwidth]{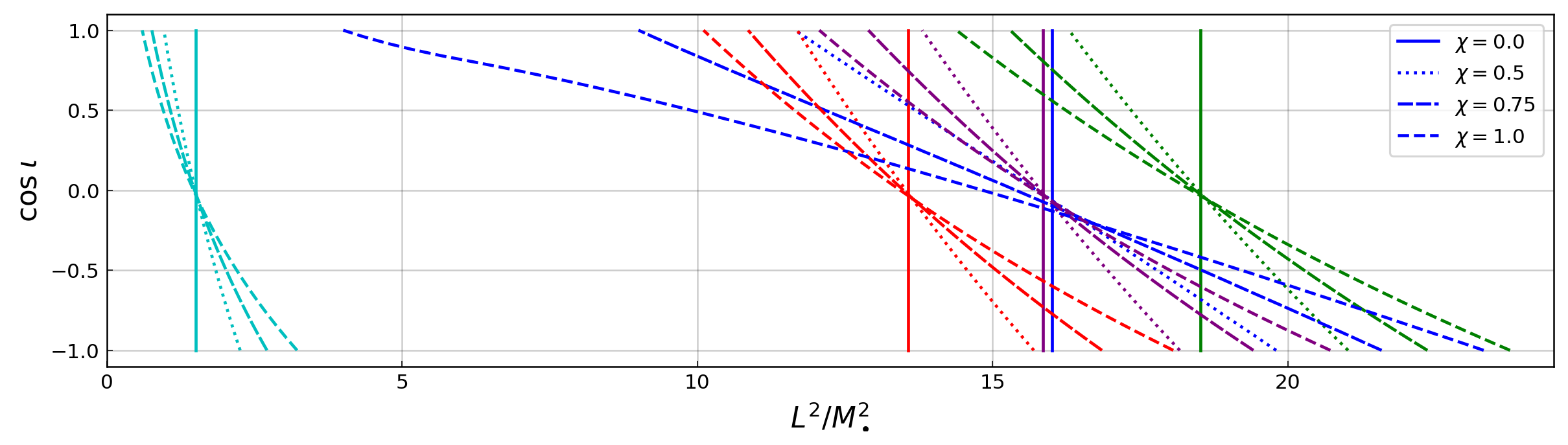}
    \includegraphics[width=\textwidth]{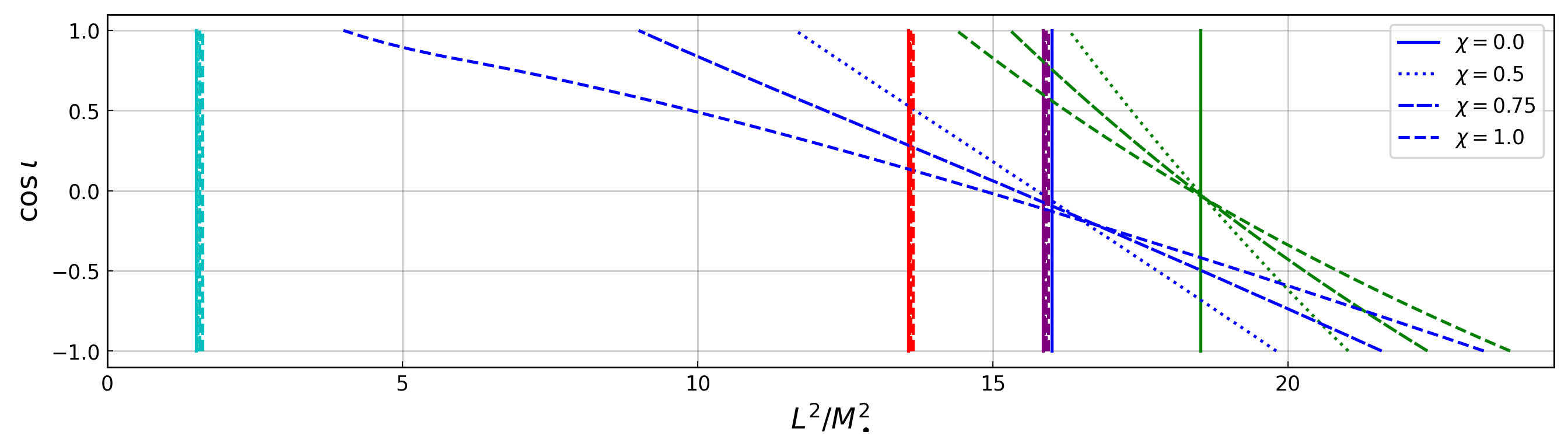}
    \caption{
    The angular-momentum thresholds for tidal disruption $L_d$ (green), capture by the event horizon $L_{\rm cap}$ (blue), and emptiness of the loss cone $L_0$ ($\mathcal{E}^\ast = 1$ --- cyan, $\mathcal{E}^\ast = 5$ --- red, $\mathcal{E}^\ast = 10$ --- purple) as functions of the cosine of the orbital inclination $\iota$.  The SMBH mass is $M_\bullet = 10^7 M_\odot$ and the solid, dotted, dot-dashed, and dashed curves correspond to SMBH spins $\chi = 0, 0.5, 0.75$, and 1.  The upper panel assumes that stellar diffusion preserves the orbital inclination, while the bottom panel assumes this diffusion fully isotropizes the distribution function.
    }
    \label{fig: C&D with LC for 10^7}
\end{figure*}

\subsection{Relativistic Corrections}

Relativistic effects associated with the Kerr metric that describes spinning SMBHs change the Newtonian calculation presented in the previous subsection in three ways:

\begin{enumerate}

\item The SMBH spin breaks the spherical symmetry of a Newtonian point mass into axisymmetry about the spin direction.  This introduces dependence on the orbital inclination $\iota$ into the differential TDE rate:
\begin{equation} \label{E:tripdif}
\frac{d^3\dot{N}}{d\mathcal{E}^\ast d(L^2/M_\bullet^2)d(\cos\iota)} = \frac{1}{2} \left( \frac{2\pi\sigma GM_\bullet}{c} \right)^2 f(\mathcal{E}, \mathcal{R}, \iota)~.
\end{equation}

\item The dependence of the relativistic tidal tensor on the SMBH spin $\chi$ and inclination $\iota$ introduces dependence on these parameters into the upper bound $L_d(\chi, \iota)$ of the loss cone for tidal disruption as described in Sec.~\ref{SS:TD}.

\item The possibility of direct capture by the SMBH event horizon sets a lower bound $ L_{\rm cap}(\chi, \iota)$ on the range of angular momenta leading to observable TDEs as described in Sec.~\ref{SS:cap}.

\end{enumerate}
These three changes imply that the TDE rate per unit dimensionless specific binding energy will be given by
\begin{align}
\frac{d\dot{N}}{d\mathcal{E}^\ast} &\approx \frac{2\sqrt{2}\sigma^5\mathcal{E}^{\ast 1/2}}{\pi GM_\star c^2} A_{\rm FR}(\chi) \label{E:dif_FLC_rel} \\
&\approx \frac{2\sqrt{2}\sigma^5\mathcal{E}^{\ast 1/2}}{\pi GM_\star c^2} A_{\rm ER}(\chi, \mathcal{E}^\ast)~, \label{E:dif_ELC_rel}
\end{align}
in the full and empty loss-cone regime, where
\begin{align}
A_{\rm FR}(\chi) &\equiv \frac{1}{2} \int_{-1}^{1} d(\cos\iota) \int_{L_{\rm cap}^2(\chi, \iota)/M_\bullet^2}^{L_d^2(\chi, \iota)/M_\bullet^2} d(L^2/M_\bullet^2) \notag \\
&= \frac{1}{2} \int_{-1}^{1} \left[ \frac{L_d^2(\chi, \iota)}{M_\bullet^2} -  \frac{L_{\rm cap}^2(\chi, \iota)}{M_\bullet^2} \right] d(\cos\iota) \label{E:A_FR} \\
A_{\rm ER}(\chi, \mathcal{E}^\ast) &\equiv \frac{1}{2} \int_{-1}^{1} d(\cos\iota) \int_{{\rm min}\{L_{0}^2(\chi, \iota),\,L_{\rm cap}^2(\chi, \iota)\}/M_\bullet^2}^{L_d^2(\chi, \iota)/M_\bullet^2} \notag \\
&\quad \quad \times \frac{f_{\rm ELC}(\mathcal{E}, \mathcal{R}, \iota)}{f_{\rm FLC}(\mathcal{E})}\, d(L^2/M_\bullet^2)~. \label{E:A_ER}
\end{align}
To properly evaluate the anisotropic distribution function $f_{\rm ELC}(\mathcal{E}, \mathcal{R}, \iota)$ and the lower limit $L_{0}^2(\chi, \iota)$ appearing in Eq.~(\ref{E:A_ER}), one would need to self-consistently derive coefficients for and then solve the orbit-averaged Fokker-Planck equation subject to the anisotropic boundary conditions set by the loss-cone boundary
\begin{equation} \label{E:LCB}
L_{\rm lc}(\chi, \iota) \equiv {\rm max}\{L_{\rm cap}(\chi, \iota), L_d(\chi, \iota)\}~.
\end{equation}
Performing this elaborate calculation would be a worthy endeavor but is beyond the scope of what we intend in this paper.

Instead, we will adopt two extreme simplifying approximations:
\begin{enumerate}

\item \emph{Inclination-preserving (IP):} Stellar diffusion preserves the orbital inclination $\iota$.  The dimensionless angular momentum $\mathcal{R}_{\rm 0,IP}(\chi, \iota)$ of the lowest occupied orbit is given by Eq.~(\ref{E:R_0}) with $L_{\rm lc}(\chi, \iota)$ given by Eq.~(\ref{E:LCB}). Inserting this result into Eq.~(\ref{E:f_elc}) yields the inclination-preserving phase-space distribution function:
\begin{equation} \label{E:f_IP}
f_{\rm IP}(\mathcal{E}, \mathcal{R}, \chi, \iota) = \frac{\ln[\mathcal{R}/\mathcal{R}_{\rm 0,IP}(\mathcal{E}, \chi, \iota)]}{\ln[1/\mathcal{R}_{\rm 0,IP}(\mathcal{E}, \chi, \iota)]}f_{\rm FLC}(\mathcal{E})~.
\end{equation}

\item \emph{Isotropizing (ISO):} Stellar diffusion isotropizes the orbital inclination $\iota$. We approximate this effect by inserting the inclination-averaged loss-cone boundary
\begin{equation} \label{E:LCB_ISO}
\bar{L}_{\rm lc}^2(\chi) = \frac{1}{2} \int_{-1}^1 L_{\rm lc}^2(\chi, \iota)\,d(\cos\iota)
\end{equation}
into Eq.~(\ref{E:R_0}) to yields a lowest occupied orbit $\mathcal{R}_{\rm 0,ISO}(\chi)$ and phase-space distribution function
\begin{equation} \label{E:f_ISO}
f_{\rm ISO}(\mathcal{E}, \mathcal{R}, \chi)
= \frac{\ln[\mathcal{R}/\mathcal{R}_{\rm 0,ISO}(\mathcal{E}, \chi)]}{\ln[1/\mathcal{R}_{\rm 0,ISO}(\mathcal{E}, \chi)]} f_{\rm FLC}(\mathcal{E})
\end{equation}
that are isotropic, i.e. independent of $\iota$.  It is important to note that the loss-cone boundary maintains its inclination dependence; the averaged result of Eq.~(\ref{E:LCB_ISO}) is only used to determine the isotropic distribution function of Eq.~(\ref{E:f_ISO}).

\end{enumerate}

Stellar diffusion in angular-momentum space due to small-angle gravitational deflections by stars in the host galaxy tends to isotropize the distribution function, in much the same way that particle diffusion in coordinate space isotropizes particle concentrations.  The inclination-preserving (IP) and isotropizing (ISO) approximations are extreme in that reality will be somewhere in between the two limits in which diffusion is completely ineffective (IP) or effective (ISO) in eliminating the inclination dependence of the distribution function.
The IP approximation should be accurate in the limit of large SMBH masses $M_\bullet$ and binding energies $\mathcal{E}^\ast$ where stellar diffusion is inefficient on the orbital timescale, while the ISO approximation should be accurate in the opposite limit of small SMBH masses $M_\bullet$ and binding energies $\mathcal{E}^\ast$ where stellar diffusion can wash out the mild anisotropies in the distribution function generated by weak relativistic effects.

Fig.~\ref{fig: C&D with LC for 10^7} illustrates the effect of these two approximations on the angular momentum threshold $L_0$ below which the distribution function $f_{\rm ELC}(\mathcal{E}, \mathcal{R}, \iota)$ vanishes.  The green and blue curves, showing the disruption and capture thresholds $L_d$ and $L_{\rm cap}$ respectively, are the same in both panels.  However, the cyan, red, and purple curves, corresponding to the thresholds $L_0$ below which the loss cone is empty for dimensionless specific binding energies $\mathcal{E}^\ast = 1, 5,$ and 10, are proportional to $L_{\rm lc}$ in the inclination-preserving approximation shown in the upper panel but are vertical in the isotropizing approximation shown in the bottom panel.  The increased area between the $L_0$ and $L_d$ curves in the latter case will lead to comparative bias towards retrograde ($\cos\iota < 0$) orbits for observable TDEs.  We explore this bias in Sec.~\ref{S:Distribution of Inclinations} below; the large differences between the inclination-preserving and isotropizing approximations motivate a more sophisticated treatment of orbital inclinations in future work.

\begin{figure}[t!]
\includegraphics[width=\linewidth]{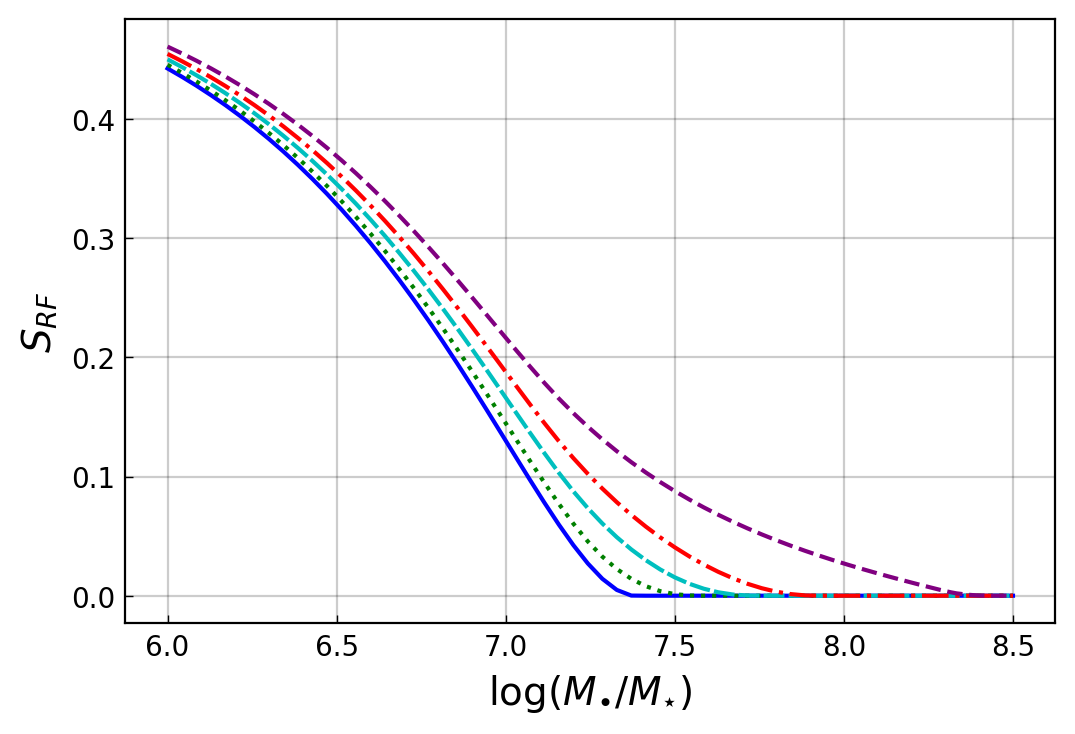}
\caption{
The factor $S_{\rm RF}$ by which the TDE rate is suppressed due to direct capture by the event horizon as a function of SMBH mass $M_\bullet$.  The solid blue, dotted green, long-dashed cyan, dot-dashed red, and short-dashed purple curves correspond to SMBH spins of $\chi = 0, 0.5, 0.75, 0.9,$ and 1.
}
\label{fig:S_RF vs Mass}
\end{figure}

\begin{figure*}[t!]
\begin{subfigure}[b]{0.49\textwidth}
\includegraphics[width=\linewidth]{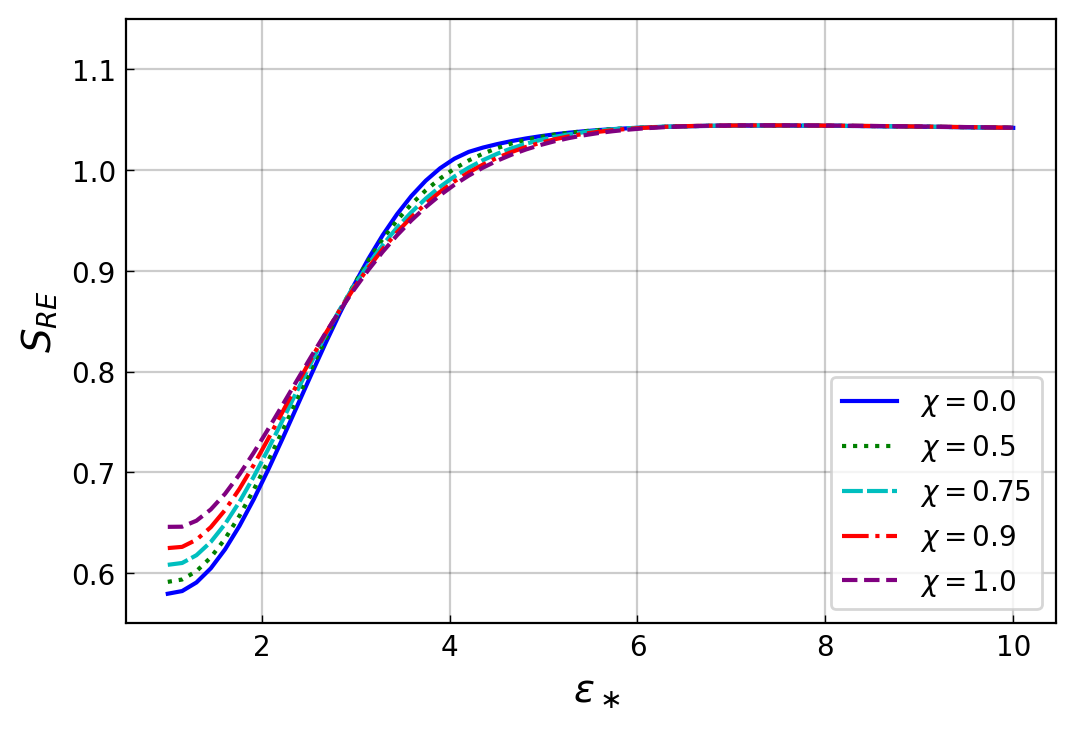}
\end{subfigure}
\begin{subfigure}[b]{0.49\textwidth}
\includegraphics[width=\linewidth]{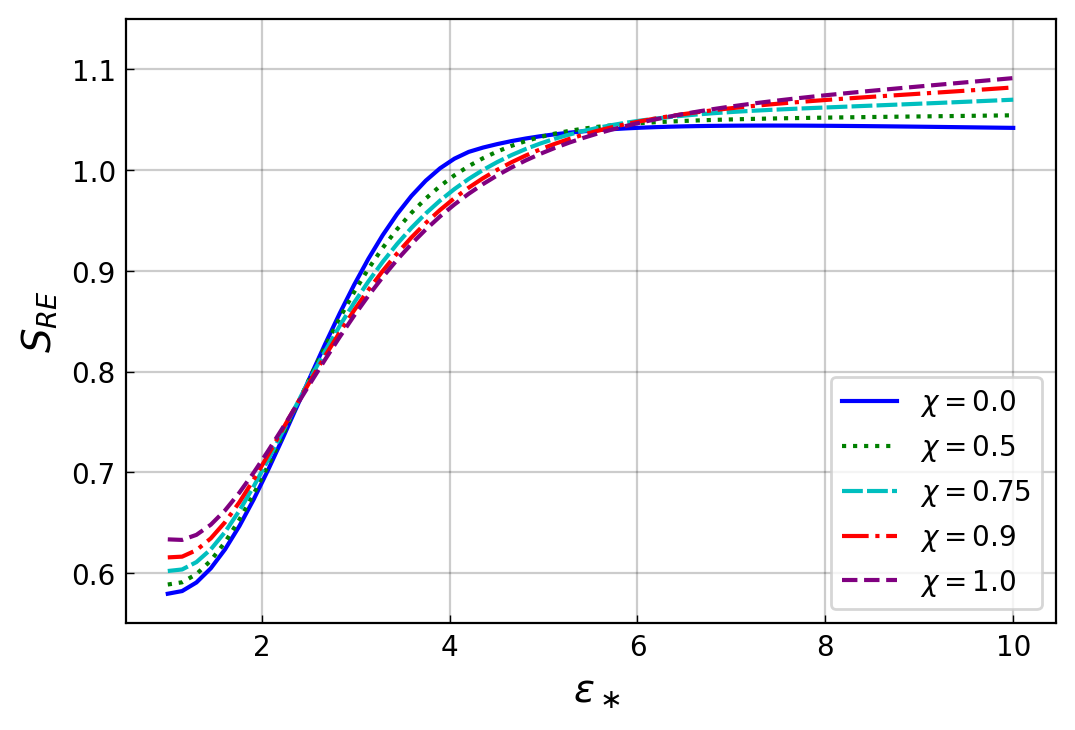}
\end{subfigure}
\caption{
The ratio $S_{\rm RE} \equiv A_{\rm ER}/A_{\rm EN}$ between the TDE rate in relativity and Newtonian gravity for loss cones filled by stellar diffusion as a function of dimensionless specific binding energy $\mathcal{E}^\ast$.  The SMBH mass is $M_\bullet = 10^7 M_\odot$ and the line styles and colors identify the SMBH spin as indicated in the caption to Fig.~\ref{fig:S_RF vs Mass}.  The left (right) panel corresponds to the limit that stellar diffusion is inclination-preserving (isotropizing).
}
\label{fig: S_RE vs epsilon}
\end{figure*}

In Fig.~\ref{fig:S_RF vs Mass}, we plot the ratio $S_{\rm RF} \equiv A_{\rm FR}(\chi)/A_{\rm FN}$ of the TDE rate in general relativity to that in Newtonian gravity in the full-loss-cone limit as a function of SMBH mass $M_\bullet$.  We see that direct capture by the event horizon suppresses the TDE rate by over 50\% even at SMBH masses as low as $10^6 M_\odot$.  The asymmetry in the inclination dependence of the tidal disruption and capture thresholds shown in Fig.~\ref{fig: C&D Angular momenta} implies that SMBH spin mildly enhances the TDE rate at low SMBH masses and increases the maximum mass capable of observable tidal disruption from $10^{7.38}M_\odot$ for Schwarzschild ($\chi = 0$) SMBHs \cite{ServinKesden2017} to above $10^{8.4} M_\odot$ for maximally spinning ($\chi = 1$) Kerr SMBHs.

\begin{figure*}[t!]
\begin{subfigure}[b]{0.49\textwidth}
\includegraphics[width=\linewidth]{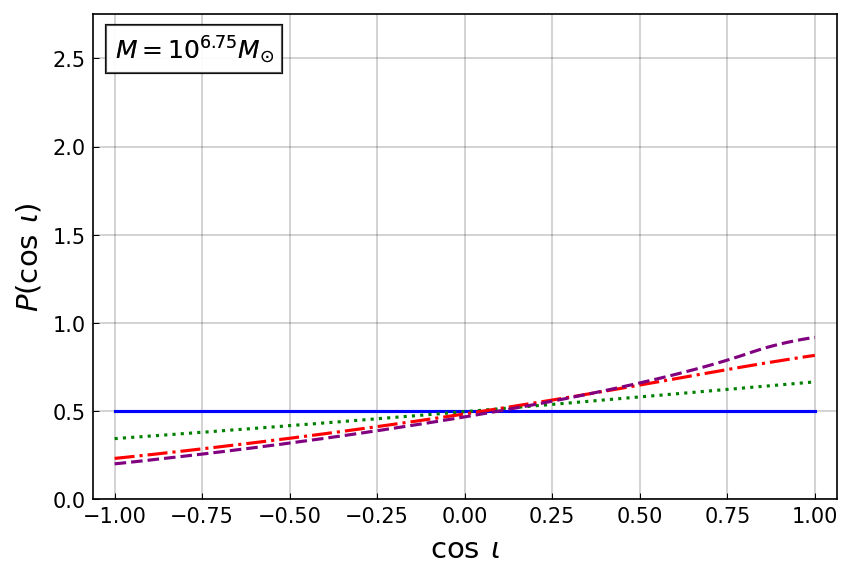}
\end{subfigure}
\begin{subfigure}[b]{0.49\textwidth}
\includegraphics[width=\linewidth]{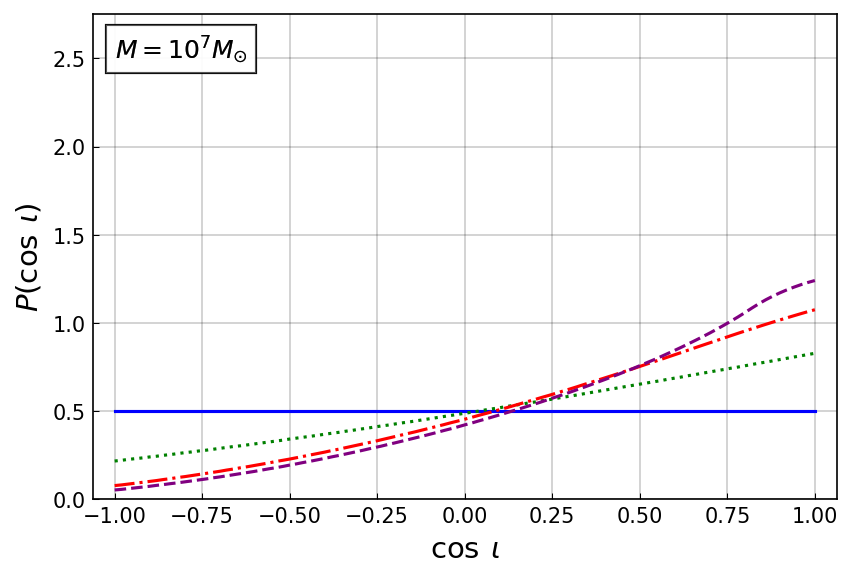}
\end{subfigure}
\begin{subfigure}[b]{0.49\textwidth}
\includegraphics[width=\linewidth]{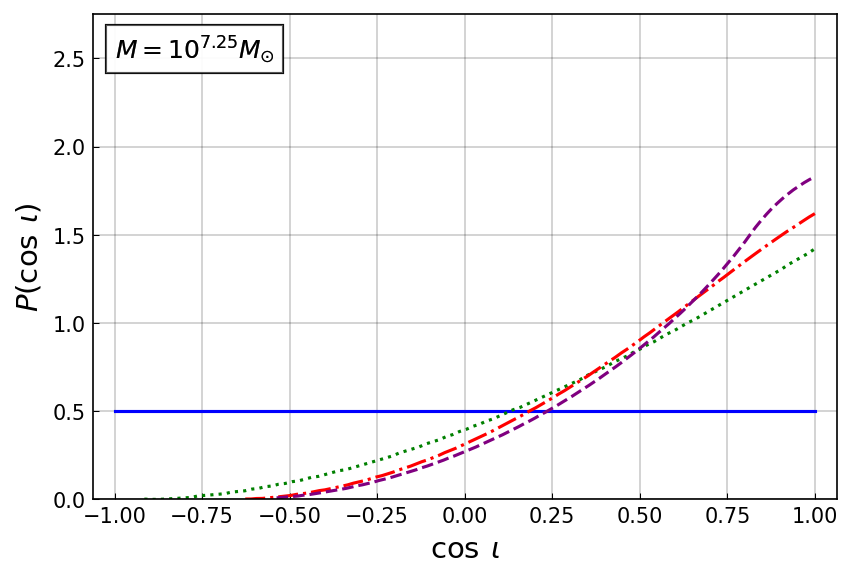}
\end{subfigure}
\begin{subfigure}[b]{0.49\textwidth}
\includegraphics[width=\linewidth]{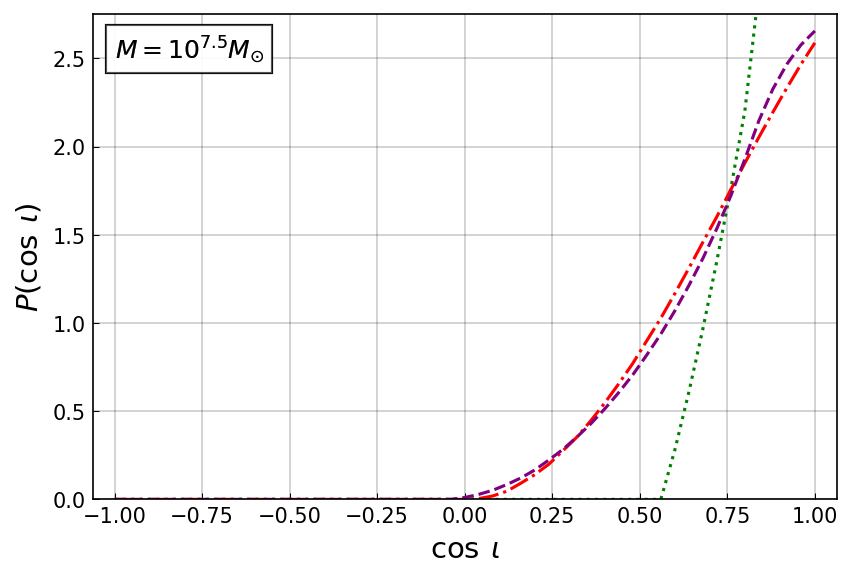}
\end{subfigure}
\caption{Probability distribution functions for TDEs inclinations for full loss cones.  As in Fig.~\ref{fig: C&D Angular momenta}, top left, top right, bottom left, and bottom right panels correspond to SMBH masses $M_\bullet = 10^{6.75}, 10^7, 10^{7.25},$ and $10^{7.5}\,M_\odot$ respectively, and the solid blue, dotted green, dot-dashed red, and dashed purple curves correspond to SMBH spins $\chi = 0, 0.5, 0.9,$ and 1.  In the bottom right panel, the Schwarzschild case ($\chi = 0$) is not shown because direct capture by the event horizon drives the TDE rate to zero for SMBH masses above $10^{7.39} M_\odot$.
}
\label{F:IncFLC}
\end{figure*}

In Fig.~\ref{fig: S_RE vs epsilon}, we examine how relativity affects TDE rates when stellar diffusion determines the occupancy of the loss cone.  The ratio $S_{\rm RE}(\chi, \mathcal{E}^\ast) \equiv A_{\rm ER}(\chi, \mathcal{E}^\ast)/A_{\rm EN}(\mathcal{E}^\ast)$, where $A_{\rm ER}$ and $A_{\rm EN}$ are given by Eqs.~(\ref{E:A_ER}) and (\ref{E:A_EN}) respectively, quantifies how relativity affects the differential TDE rate $d\dot{N}/d\mathcal{E}^\ast$ as a function of dimensionless specific binding energy $\mathcal{E}^\ast$.  At small values of $\mathcal{E}^\ast$, the long orbital periods imply that stellar diffusion has time to refill the loss cone.  The angular momentum $L_0(\mathcal{E}^\ast)$ below which the distribution function  $f_{\rm ELC}(\mathcal{E}^\ast, L, \iota)$ vanishes is well below the capture threshold $L_{\rm cap}$ as can be seen for the cyan curves ($\mathcal{E}^\ast = 1$) in Fig.~\ref{fig: C&D with LC for 10^7}.  Stars can thus be captured by the event horizon and the ratio $S_{\rm RE}$ is below unity.  The asymmetry of capture with respect to orbital inclination implies that this suppression is weaker for highly spinning SMBHs for which stars can preferentially avoid capture on prograde ($\cos\iota > 0$) orbits.  As $\mathcal{E}^\ast$ increases, loss-cone repopulation becomes inefficient and $L_0(\mathcal{E}^\ast)$ increases as can be seen for the red and purple curves ($\mathcal{E}^\ast = 5$ and 10) in Fig.~\ref{fig: C&D with LC for 10^7}.  As $L_0(\mathcal{E}^\ast)$ approaches $L_{\rm cap}$, direct capture is highly suppressed, and $S_{\rm RE}$ rises above unity because the stronger tidal forces (larger values of the disruption threshold $L_d$) in relativity compared to Newtonian gravity.  This is expressed mathematically by the wider limits of integration in Eq.~(\ref{E:A_ER}) than in Eq.~(\ref{E:A_EN}), equivalent to more area between the $L_0$ and $L_d$ curves in Fig.~\ref{fig: C&D with LC for 10^7}.

The largest difference between the inclination-preserving and isotropizing approximations shown in the left and right panels of Fig.~\ref{fig: S_RE vs epsilon} occurs for highly bound orbits ($\mathcal{E}^\ast > 6$).  For the SMBH mass $M_\bullet = 10^7 M_\odot$ shown in Figs.~\ref{fig: C&D with LC for 10^7} and \ref{fig: S_RE vs epsilon}, $L_d > L_{\rm cap}$ for all SMBH spins and inclinations implying that direct capture is highly suppressed for $\mathcal{E}^\ast > 6$.  In the inclination-preserving case, the $L_0(\chi, \iota)$ curves are nearly parallel to the disruption curves $L_d(\chi, \iota)$.  This implies that the area between these curves, and thus the integral $A_{\rm ER}$ and the TDE rate, is nearly independent of their spin-dependent slopes.  However, in the isotropizing case, the $L_0$ curves are independent of the inclination $\iota$.  Greater SMBH spin increases the area between the $L_0(\chi, \iota)$ and $L_d(\chi, \iota)$ curves and the TDE rate for retrograde ($\cos\iota < 0$) orbits without an accompanying suppression in the TDE rate on prograde ($\cos\iota > 0$) orbits, which is already near zero and cannot fall further.

\section{Distribution of TDE inclinations} \label{S:Distribution of Inclinations}

The distribution of TDE inclinations depends on a subtle interplay between the inclination dependencies of the disrpution threshold $L_d$, the direct-capture threshold $L_{\rm cap}$, and the refilling of the loss cone in response to these boundary conditions in phase space.  We begin by discussing this distribution for the full loss cone that applies in the limit $M_\bullet \to 0$, then generalize to steady-state TDE rates that apply to finite SMBH masses.

\subsection{Full Loss Cone}

The probability distribution function for TDE inclinations
\begin{equation} \label{E:totIDF}
P(\cos\iota) \equiv \frac{d(\ln\dot{N})}{d(\cos\iota)}
\end{equation}
is obtained by integrating the differential TDE rate of Eq.~(\ref{E:tripdif}) over specific energy $\mathcal{E}$ and squared specific angular momentum $L^2$ over the domain $\mathcal{E} > 0$ and $L_{\rm cap}^2(\chi,\iota) < L^2 < L_d^2(\chi,\iota)$.

Doing so for the full loss cone in which the phase-space distribution function $f_{\rm FLC}$ is given by Eq.~(\ref{E:f_FLC}) yields the inclination distribution functions $P(\cos\iota)$ shown in Fig.~\ref{F:IncFLC}.  These functions are proportional to the distance between the capture and disruption curves $L_{\rm cap}^2(\chi,\iota)$ and $L_d^2(\chi,\iota)$ in Fig.~\ref{fig: C&D Angular momenta}.  Because direct capture has a stronger retrograde bias than tidal disruption, the inclination distribution of stars that are disrupted but avoid direct capture has a prograde bias that increases with SMBH mass as relativistic effects become more significant.  For $M_\bullet = 10^{7.5} M_\odot$, direct capture has fully suppressed TDEs for $\chi = 0$ and $0.5$, and only prograde TDEs ($\cos\iota > 0$) survive for $\chi = 0.9$ and $1$, albeit at highly suppressed rates.

\begin{figure*}[t!]
\begin{subfigure}[b]{0.49\textwidth}
\includegraphics[width=\linewidth]{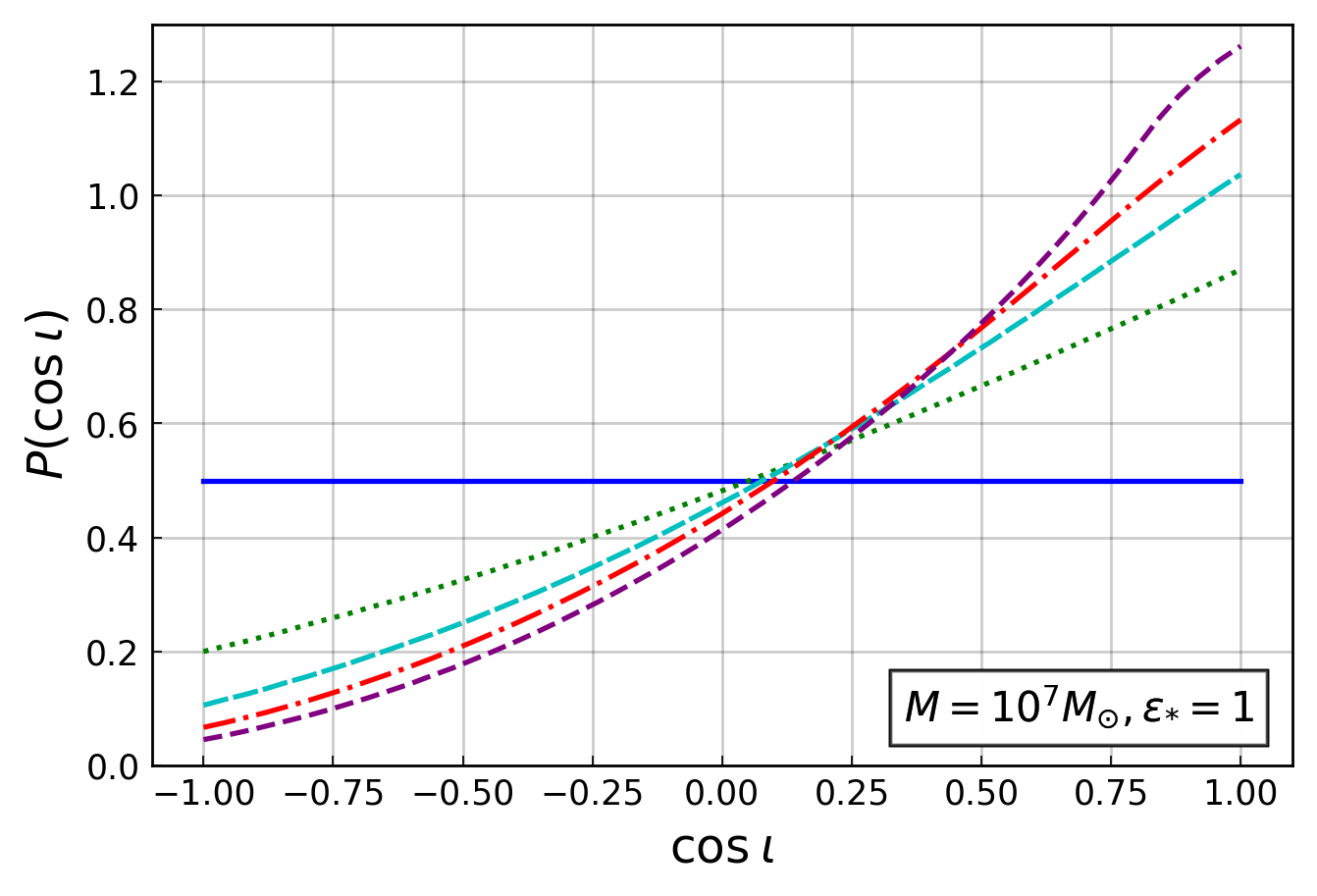}
\end{subfigure}
\begin{subfigure}[b]{0.49\textwidth}
\includegraphics[width=\linewidth]{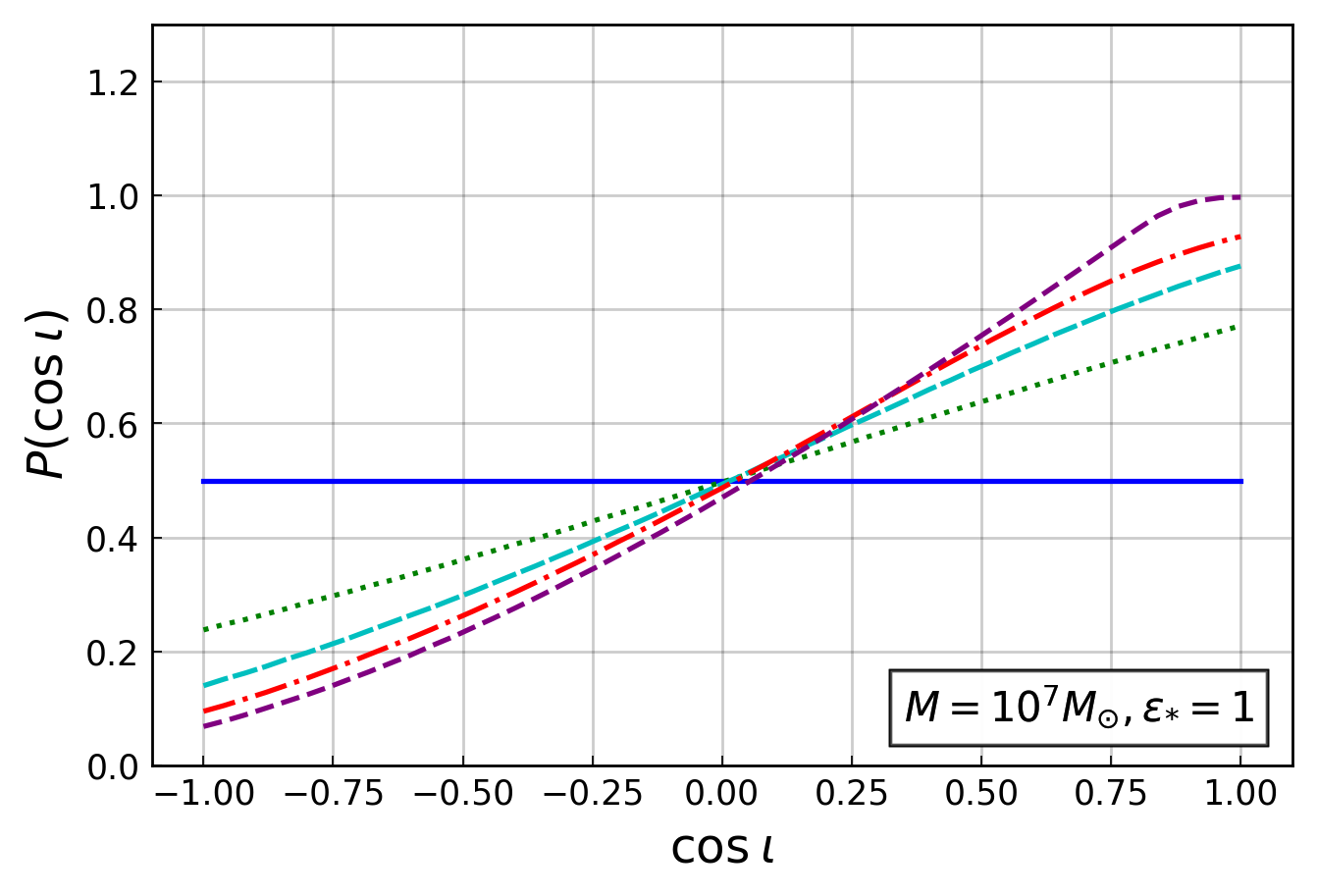}
\end{subfigure}
\begin{subfigure}[b]{0.49\textwidth}
\includegraphics[width=\linewidth]{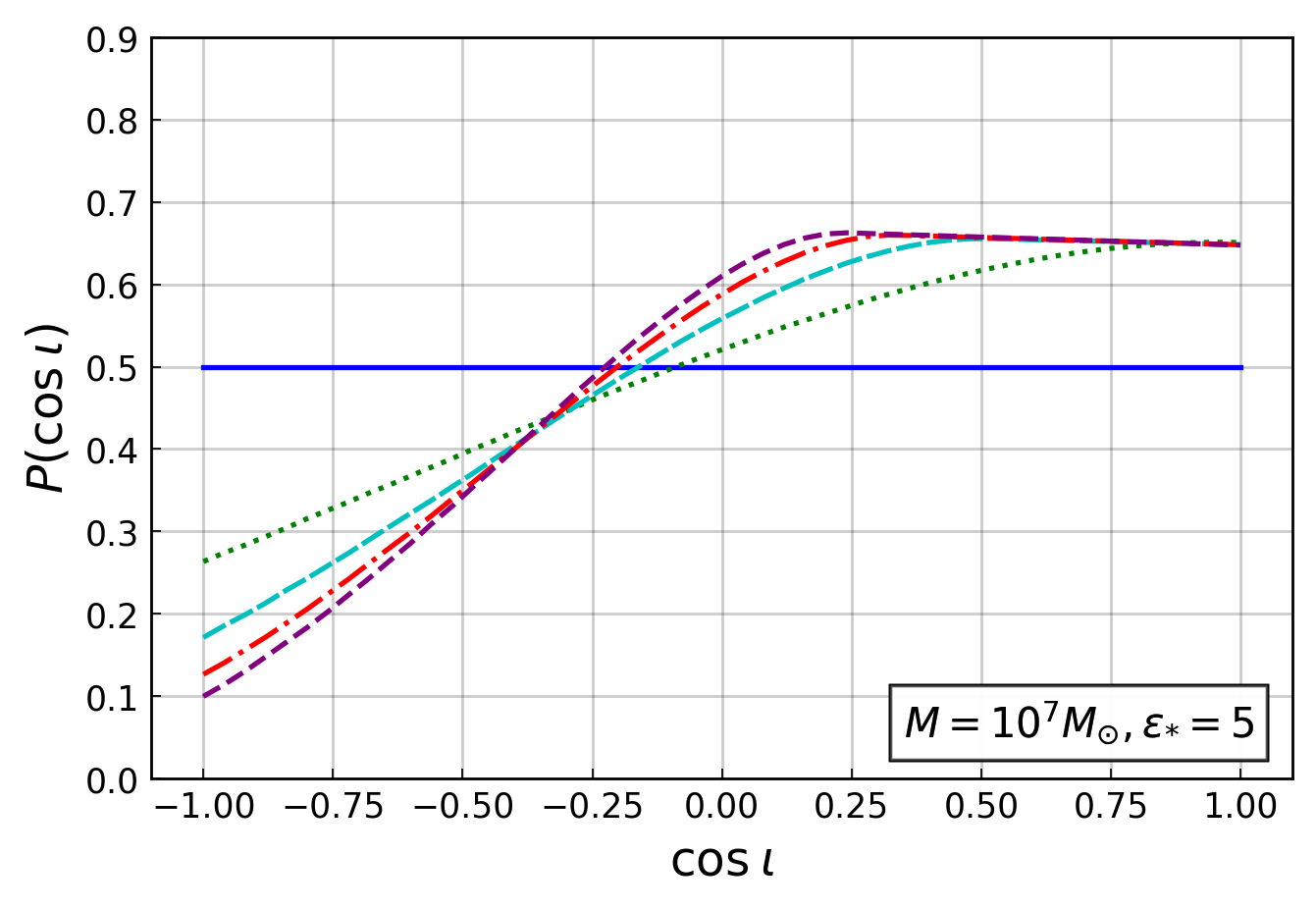}
\end{subfigure}
\begin{subfigure}[b]{0.49\textwidth}
\includegraphics[width=\linewidth]{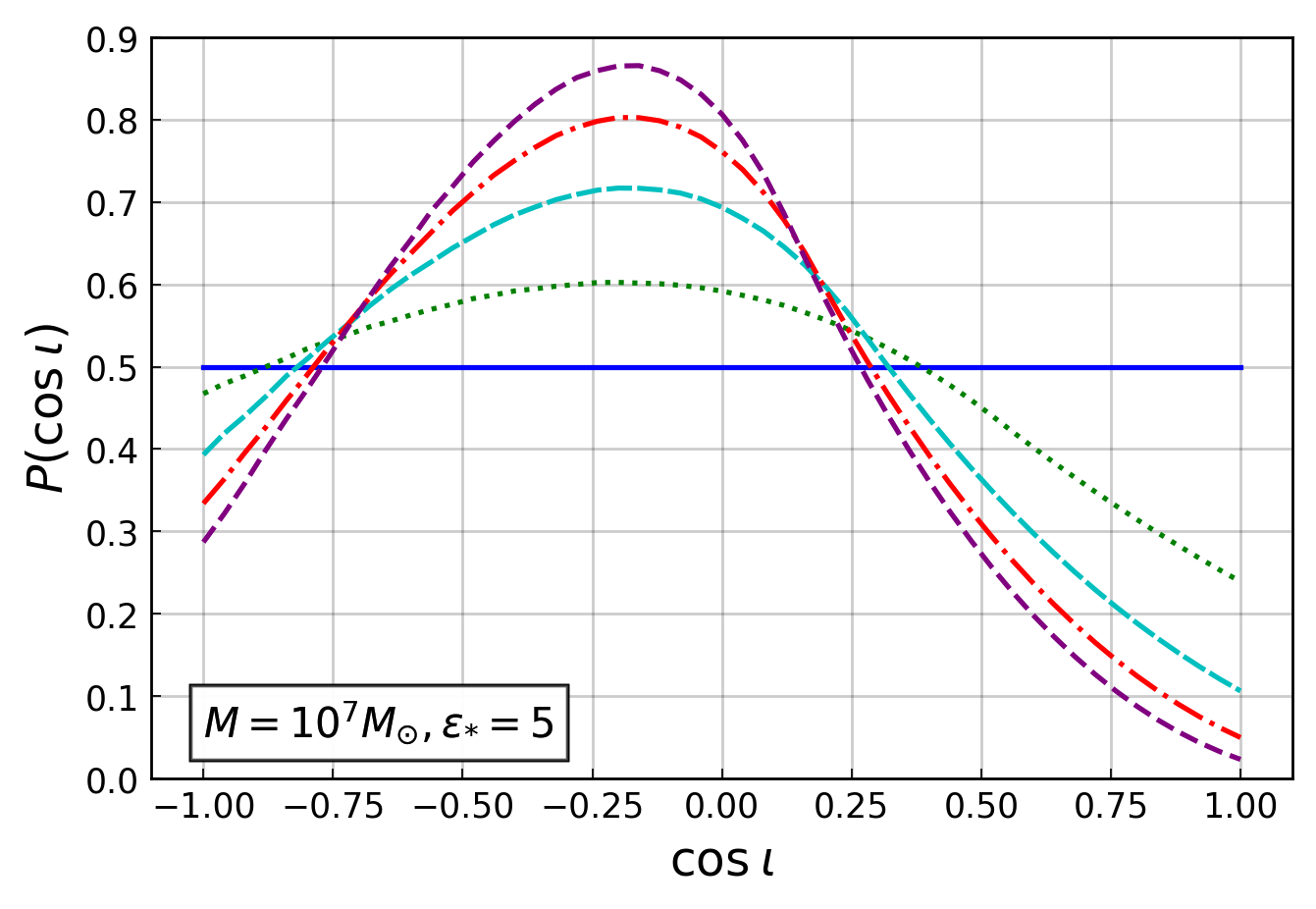}
\end{subfigure}
\begin{subfigure}[b]{0.49\textwidth}
\includegraphics[width=\linewidth]{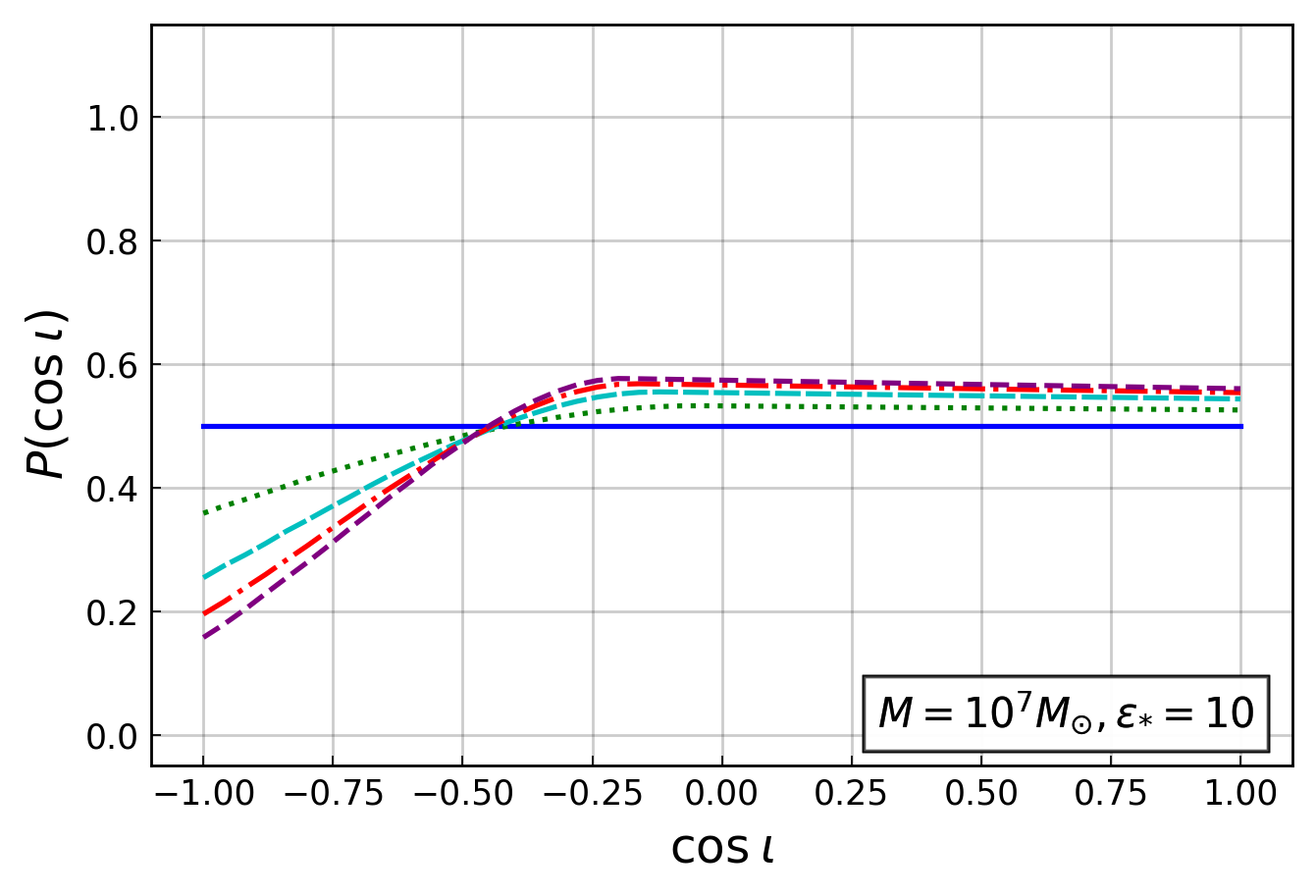}
\end{subfigure}
\begin{subfigure}[b]{0.49\textwidth}
\includegraphics[width=\linewidth]{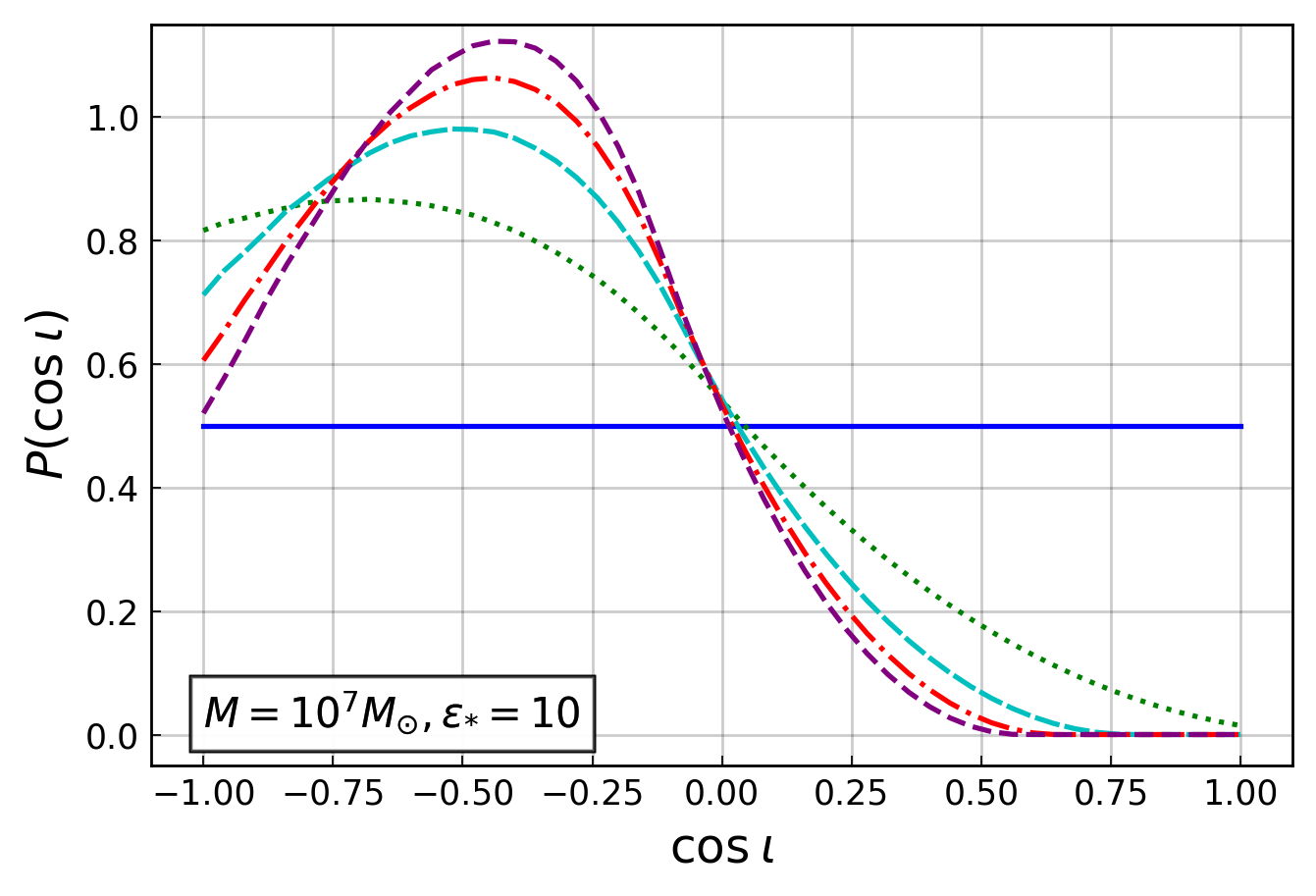}
\end{subfigure}
\caption{
Probability distribution functions for TDE inclinations assuming that the loss cone is refilled at the steady-state rate by inclination-preserving stellar scattering (left panels) or isotropizing scattering (right panels).  The SMBH mass is $M_\bullet = 10^{7}M_\odot$ and top, middle, and bottom panels correspond to dimensionless specific energies $\mathcal{E}^\ast = 1$, $5$, and $10$ respectively.  The solid blue, dotted green, long-dashed cyan, dot-dashed red, and short-dashed purple curves correspond to SMBH spins $\chi = 0, 0.5, 0.75, 0.9,$ and 1.
}
\label{fig: PDFs for Loss Cone, Energy}
\end{figure*}

\begin{figure*}[t!]
\begin{subfigure}[b]{0.49\textwidth}
\includegraphics[width=\linewidth]{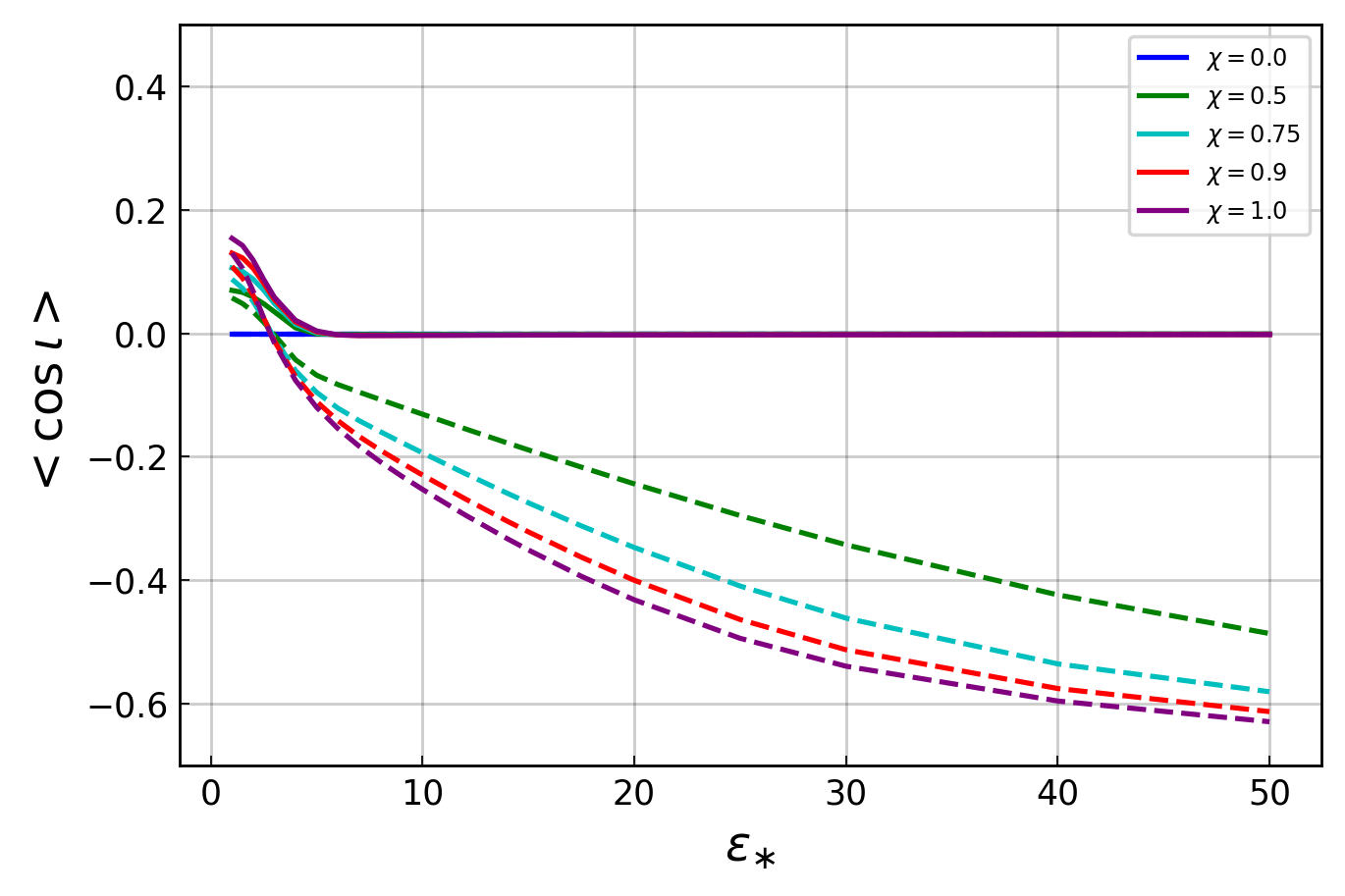}
\end{subfigure}
\begin{subfigure}[b]{0.49\textwidth}
\includegraphics[width=\linewidth]{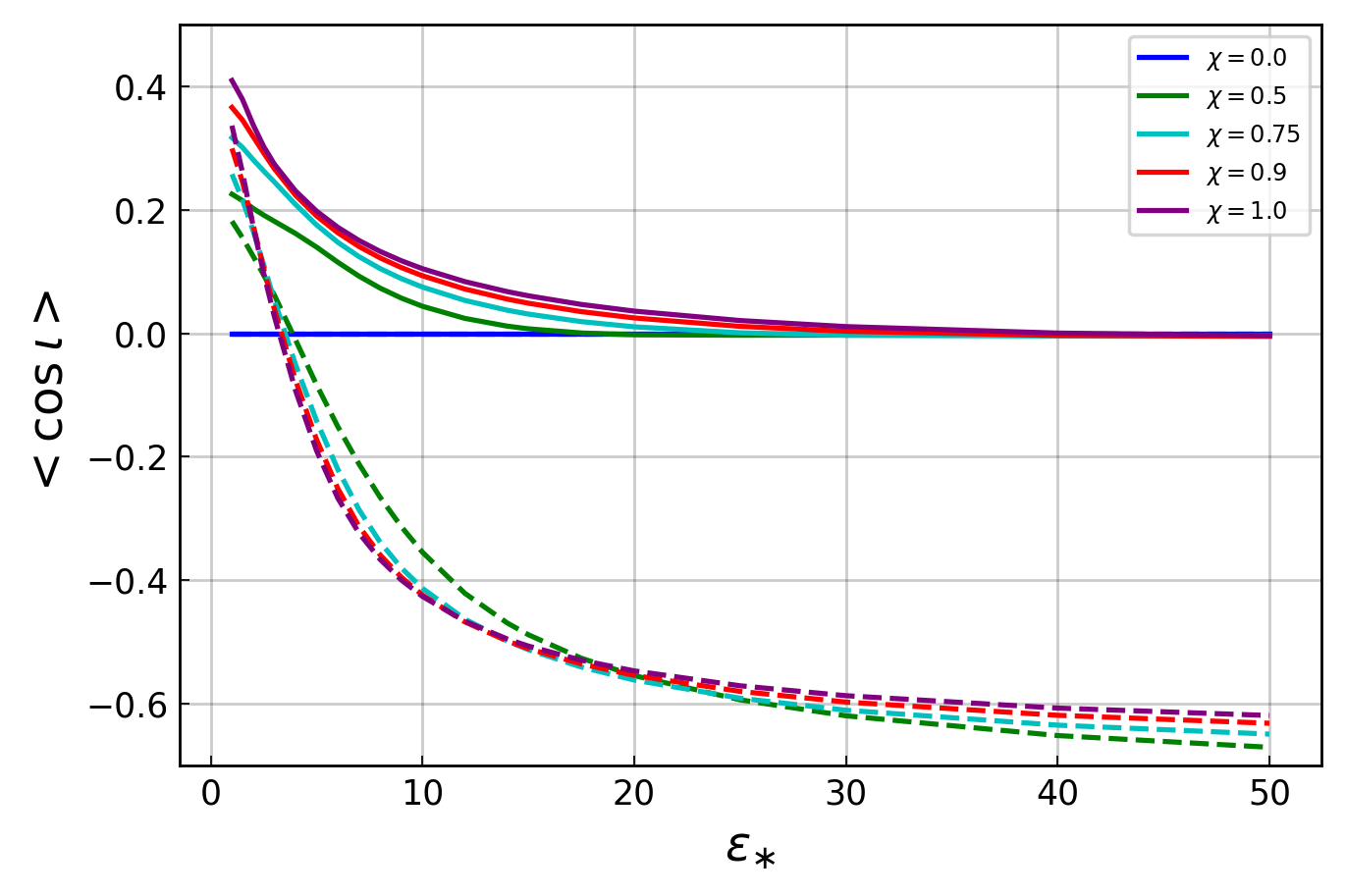}
\end{subfigure}
\caption{Mean inclination $\langle\cos\iota\rangle$ as a function of specific binding energy $\mathcal{E}^\ast$ for SMBH mass $10^{6.5} M_\odot$ (left panel) and $10^7 M_\odot$ (right panel).  The solid (dashed) lines show the limits of inclination-preserving (isotropizing) refilling of the loss cone.  The blue, green, cyan, red, and purple curves correspond to SMBH spins $\chi = 0, 0.5, 0.75, 0.9,$ and 1.
}
\label{fig: Mean cos i}
\end{figure*}

\subsection{Steady-State Loss Cone}

In the steady-state limit, the occupancy of the loss cone is determined by equilibrium between the rate at which it is emptied by tidal disruption and direct capture and the rate at which it is refilled by the scattering of stars with $L > L_{\rm lc}$ into the loss cone.  We consider the two extreme cases in which this stellar scattering which repopulates the loss cone either preserves the mean stellar inclination or fully isotropizes it.  The phase-space distribution functions $f_{\rm IP}$ and $f_{\rm ISO}$ are given by Eqs.~(\ref{E:f_IP}) and (\ref{E:f_ISO}) in these two limits.

In the left and right columns of Fig.~\ref{fig: PDFs for Loss Cone, Energy}, we show the inclination distribution functions in these two cases for fixed dimensionless specific stellar binding energy $\mathcal{E}^\ast$:
\begin{equation} \label{E:partIDF}
P(\cos\iota, \mathcal{E}^\ast) \propto \frac{d^2\dot{N}}{d(\cos\iota)d\mathcal{E}^\ast}~.
\end{equation}
We hold $\mathcal{E}^\ast$ fixed in the panels of this figure to illustrate the effects of these two assumptions as the loss cone becomes increasing empty at higher values of $\mathcal{E}^\ast$ (shorter orbital periods at higher binding energies provide less time to refill the loss cone).  For the lowest binding energy $\mathcal{E}^\ast = 1$ shown in the top row, the lowest occupied orbit $L_0$ is well below the loss-cone boundary $L_{\rm lc}$ in both the inclination-preserving and isotropizing limits; compare the cyan curves $L_0$ to the green and blue disruption and capture curves $L_d$ and $L_{\rm cap}$ in Fig.~\ref{fig: C&D with LC for 10^7}.  This implies that the inclination distributions $P(\cos\iota)$ are similar to that in the full loss cone shown in the top right panel of Fig.~\ref{F:IncFLC} after accounting for the different ranges on the y axis.  We see from the cyan curves of Fig.~\ref{fig: C&D with LC for 10^7} and Eqs.~(\ref{E:f_IP}) and (\ref{E:f_ISO}) that isotropization causes $L_{\rm 0,ISO}(\chi) > L_{\rm 0,IP}(\chi, \iota)$ for prograde orbits ($\cos\iota > 0$) and vice versa for retrograde orbits ($\cos\iota < 0$).  This suppresses (enhances) the differential TDE rate for prograde (retrograde) orbits in the isotropizing case, flattening the inclination distribution for $\mathcal{E}^\ast = 1$.

\begin{figure*}[t!]
\begin{subfigure}[b]{0.49\textwidth}
\includegraphics[width=\linewidth]{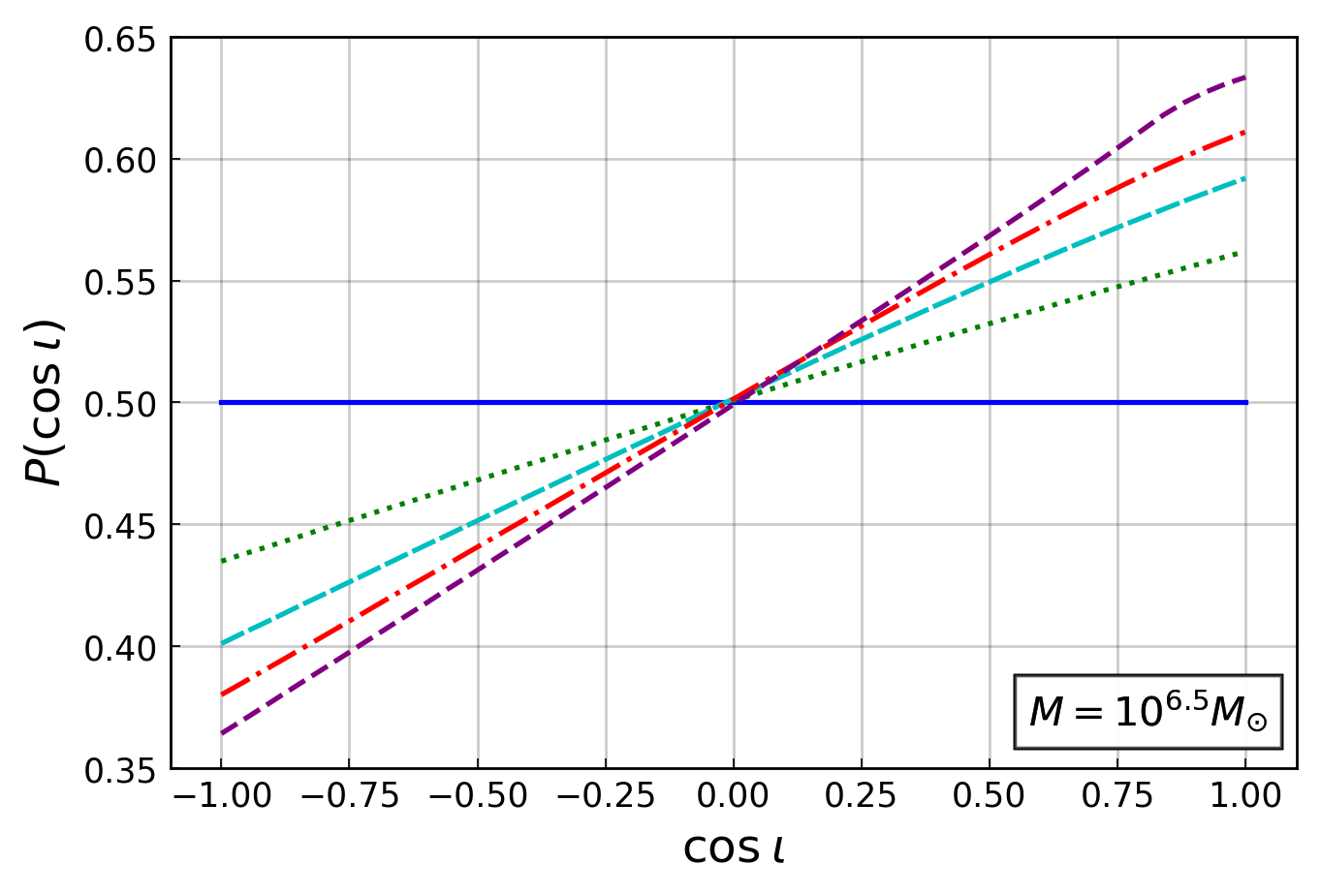}
\end{subfigure}
\begin{subfigure}[b]{0.49\textwidth}
\includegraphics[width=\linewidth]{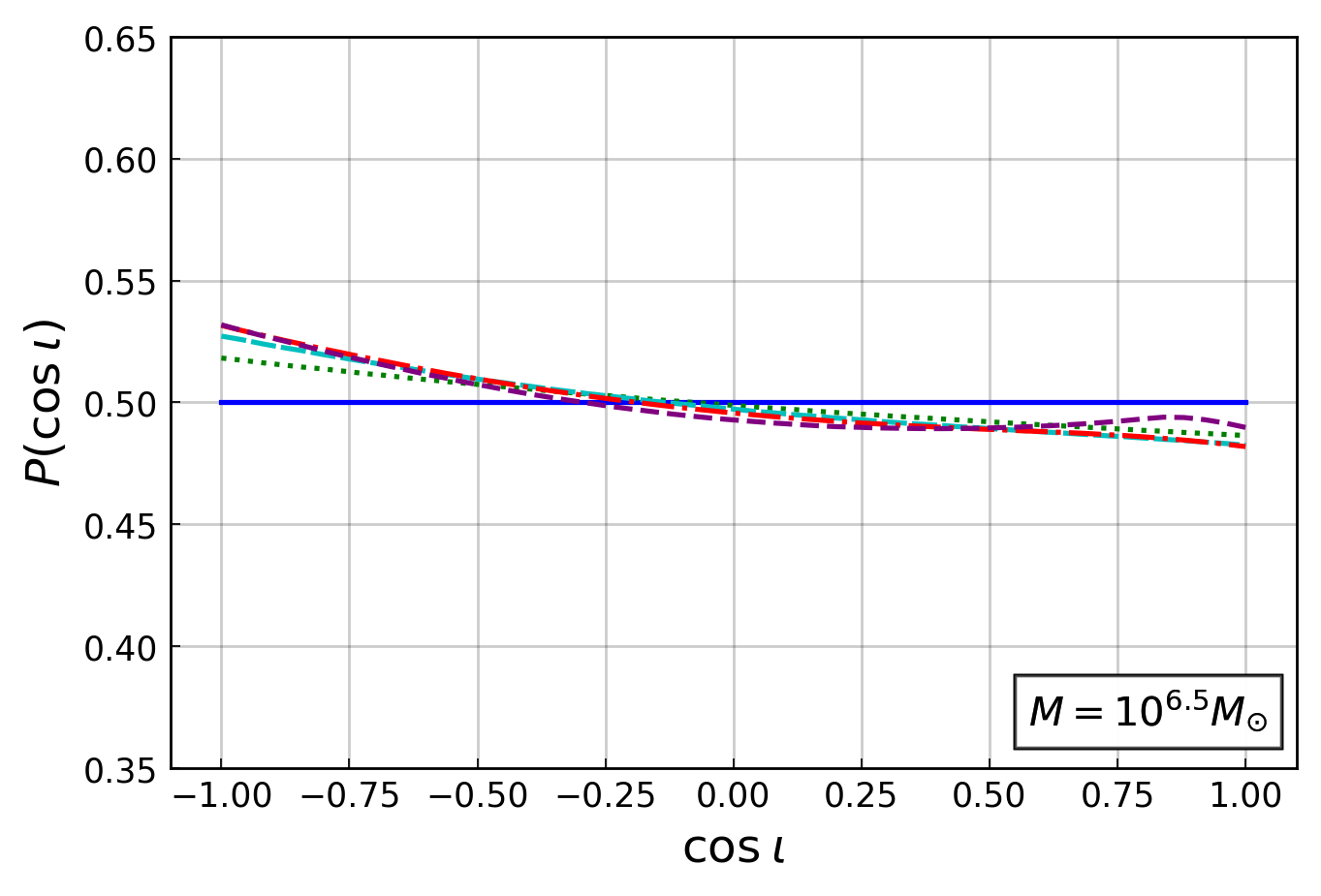}
\end{subfigure}
\begin{subfigure}[b]{0.49\textwidth}
\includegraphics[width=\linewidth]{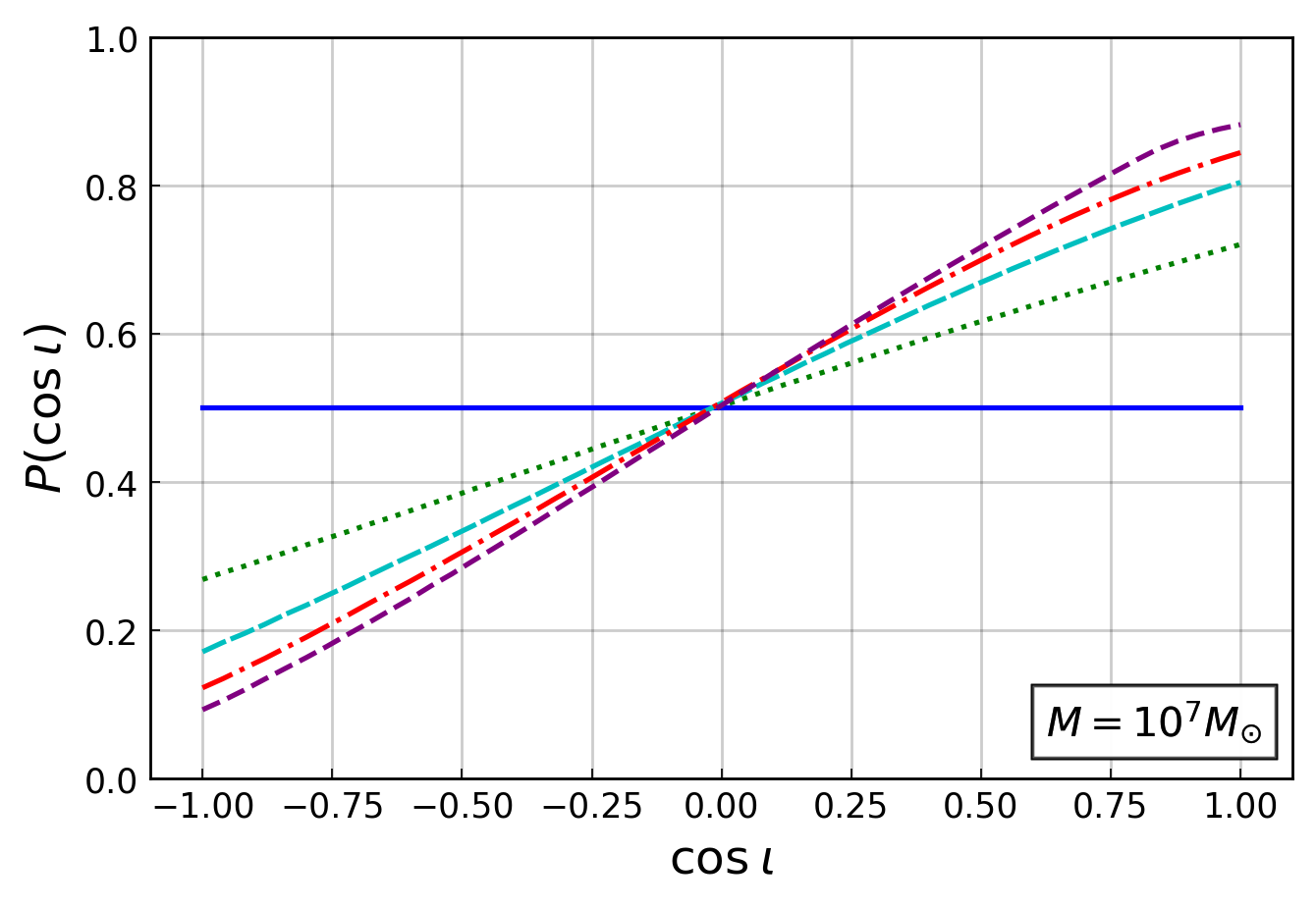}
\end{subfigure}
\begin{subfigure}[b]{0.49\textwidth}
\includegraphics[width=\linewidth]{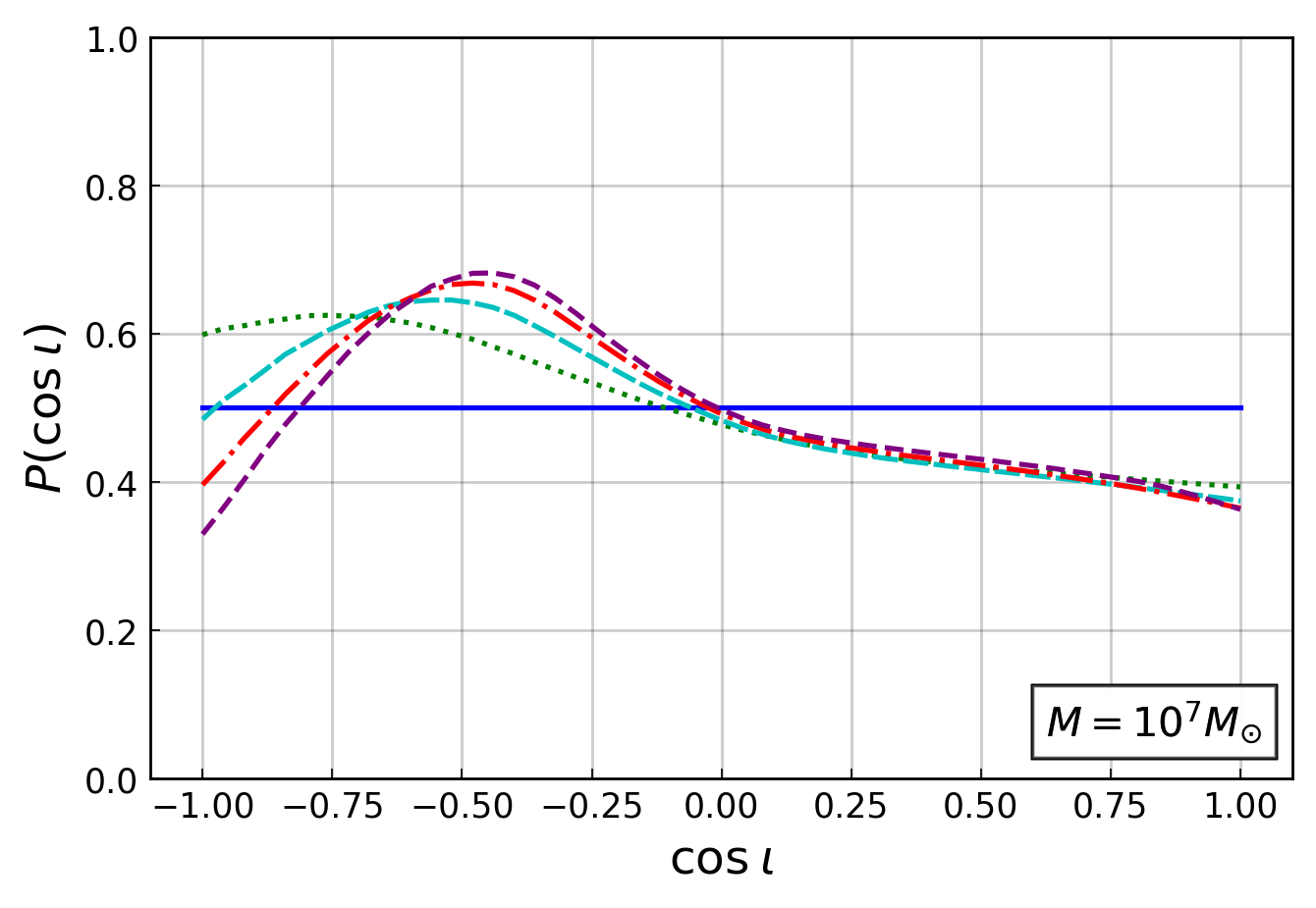}
\end{subfigure}
\caption{Probability distribution function of TDE inclinations integrated over the distribution of specific binding energies. The top (bottom) panels correspond to an SMBH mass $M_\bullet = 10^{6.5} M_\odot$ ($10^7 M_\odot$).  The left (right) panels correspond to the limit of inclination-preserving (isotropizing) refilling of loss-cone orbits.  The solid blue, dotted green, long-dashed cyan, dot-dashed red, and long-dashed purple curves correspond to SMBH spins $\chi = 0, 0.5, 0.75, 0.9,$ and 1.
}
\label{fig: PDFs for Loss Cone, Mass}
\end{figure*}

As the binding energy increases to $\mathcal{E}^\ast = 5$ and $10$ in the middle and bottom rows of Fig.~\ref{fig: PDFs for Loss Cone, Energy}, the loss cone empties and the lowest occupied orbit $L_0$ approaches the loss-cone boundary $L_{\rm lc}$ as shown by the red and purple curves in Fig.~\ref{fig: C&D with LC for 10^7}.  In the inclination-preserving limit for which $L_{\rm 0,IP}(\chi, \iota) \to L_d(\chi, \iota)$ as $\mathcal{E}^\ast \to \infty$ according to Eqs.~(\ref{E:R_0}) and (\ref{E:q}), the inclination distribution flattens for inclinations for which $L_{\rm 0,IP}(\chi, \iota) > L_{\rm cap}(\chi, \iota)$.  Capture is still able to suppress the differential TDE rate for retrograde orbits for which $L_{\rm 0,IP}(\chi, \iota) < L_{\rm cap}(\chi, \iota)$.

In the isotropizing limit shown in the right panels of Fig.~\ref{fig: PDFs for Loss Cone, Energy}, the retrograde bias caused by the stronger tidal forces on retrograde orbits ($L_d$ is a monotonically increasing function of $\iota$) can impose a corresponding retrograde bias on the inclination distribution once the isotropized lower bound to the occupied loss cone $L_{\rm 0,ISO}(\chi)$ suppresses the prograde bias in the orbits that survive direct capture, i.e. the vertical red and purple lines in the bottom panel of Fig.~\ref{fig: C&D with LC for 10^7} move to the right of the prograde portions of the blue capture curves.  We see in the middle and bottom right panels of Fig.~\ref{fig: PDFs for Loss Cone, Energy} that this effect leads to a \emph{retrograde} bias in the inclination distribution with peaks at $\cos\iota \approx -0.2$ (-0.5) for $\mathcal{E}^\ast = 5$ (10).

In Fig.~\ref{fig: Mean cos i}, we further explore the relativistic bias in TDE inclinations by calculating the mean cosine of the inclination
\begin{equation} \label{E:meancosi}
<\cos\iota>(\mathcal{E}^\ast) = \int_{-1}^{1} P(\cos\iota, \mathcal{E}^\ast)\cos\iota\,d(\cos\iota)
\end{equation}
as a function of specific binding energy $\mathcal{E}^\ast$.  We see that in the inclination-preserving limit, emptying of the loss cone at greater $\mathcal{E}^\ast$ suppresses direct capture and leaves $L_d^2(\chi, \iota) - L_{\rm 0,IP}^2(\chi, \iota)$ independent of $\iota$, flattening the inclination distribution.  These flattened distributions drive $<\cos\iota> \to 0$ for $\mathcal{E}^\ast \gtrsim 5$ (30) for $M_\bullet = 10^{6.5} M_\odot$ ($10^7 M_\odot$).  In the isotropizing limit, emptying of the loss cone at greater $\mathcal{E}^\ast$ still suppresses direct capture, however $L_d^2(\chi, \iota) - L_{\rm 0,ISO}^2(\chi)$ is now a monotonically increasing function of $\iota$ due to the retrograde bias of tidal disruption.  This implies that the mean cosine of the inclination becomes negative for $\mathcal{E}^\ast \gtrsim 2$ (3) for $M_\bullet = 10^{6.5} M_\odot$ ($10^7 M_\odot$).

One interesting feature of the isotropizing case for $M_\bullet = 10^7 M_\odot$ is that $<\cos\iota>$ becomes less negative (less retrograde bias) with increasing spin magnitude $\chi$ for $\mathcal{E}^\ast \gtrsim 20$.  This can be understood by examining the bottom panel of Fig.~\ref{fig: C&D with LC for 10^7}.  As $\mathcal{E}^\ast \to \infty$, the lowest occupied orbit $L_{\rm 0,ISO}^2(\chi)$ approaches the vertical green line corresponding to $L_d^2(\chi = 0, \iota)$ (because the near symmetry of $L_d^2(\chi, \iota)$ about $\cos\iota = 0$ produces a very weak dependence of $\bar{L}_d^2(\chi)$ on SMBH spin).  At this SMBH mass, the disruption and capture curves are sufficiently close together that capture is not fully suppressed on retrograde orbits even for $\mathcal{E}^\ast \to \infty$.  The increased capture on retrograde orbits for increasing SMBH spin $\chi$ imprints a weaker retrograde bias on the surviving TDEs, explaining this feature.

In Fig.~\ref{fig: PDFs for Loss Cone, Mass}, we integrate over the distribution of specific binding energies $\mathcal{E}^\ast$ to obtain the total inclination distribution function $P(\cos\iota)$ given by Eq.~(\ref{E:totIDF}).  Although $f_{\rm FLC} \propto (\mathcal{E}^\ast)^{1/2}$, $f_{\rm IP}, f_{\rm ISO} \to 0$ as $\mathcal{E}^\ast \to \infty$ according to Eqs.~(\ref{E:R_0}) and (\ref{E:q}) as the loss cone empties at higher binding energies (shorter periods).  This implies that the total inclination distribution function will be dominated by specific binding energies between $\mathcal{E}^\ast = 1$ and $\mathcal{E}^\ast = 5$ shown in the top and middle panels of Fig.~\ref{fig: PDFs for Loss Cone, Energy}.  We see from the left panels of Fig.~\ref{fig: PDFs for Loss Cone, Mass} that in the limit of inclination-preserving refilling of the loss cone, the stronger retrograde bias of direct capture compared to tidal disruption imposes a prograde bias on the distribution of TDE inclinations that increases with SMBH mass as relativistic effects become more significant.  However, the upper right panel shows that in the opposite extreme of isotropizing refilling of the loss cone, suppression of direct capture by the emptying loss cone has already allowed tidal disruption to impose a mild retrograde bias on the distribution of TDE inclinations by an SMBH mass of $M_\bullet = 10^{6.5} M_\odot$.  This retrograde bias becomes even more pronounced at the higher SMBH mass $M_\bullet = 10^7 M_\odot$ shown in the bottom right panel.

\begin{figure*}[t!]
\begin{subfigure}[b]{0.49\textwidth}
\includegraphics[width=\linewidth]{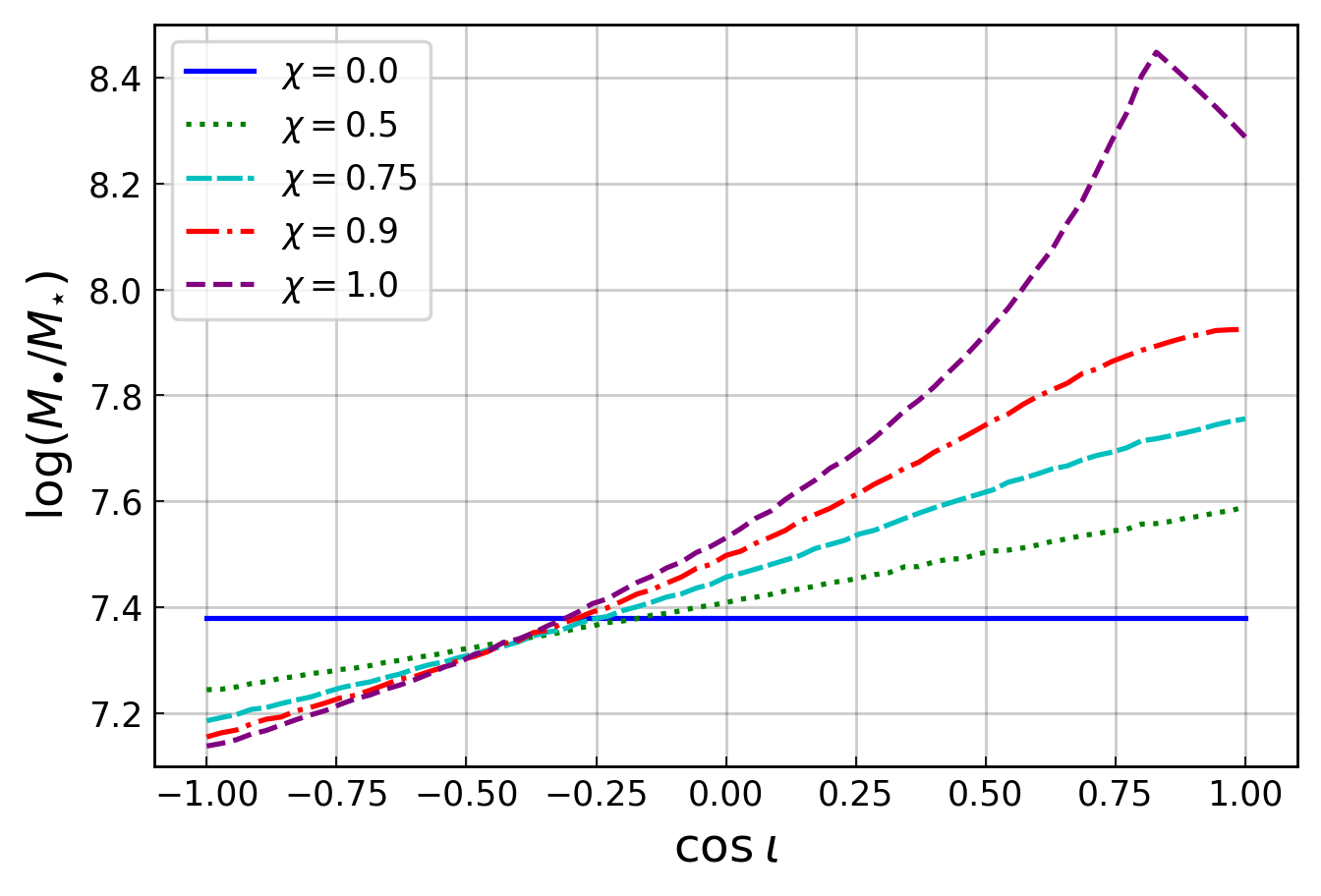}
\end{subfigure}
\begin{subfigure}[b]{0.49\textwidth}
\includegraphics[width=\linewidth]{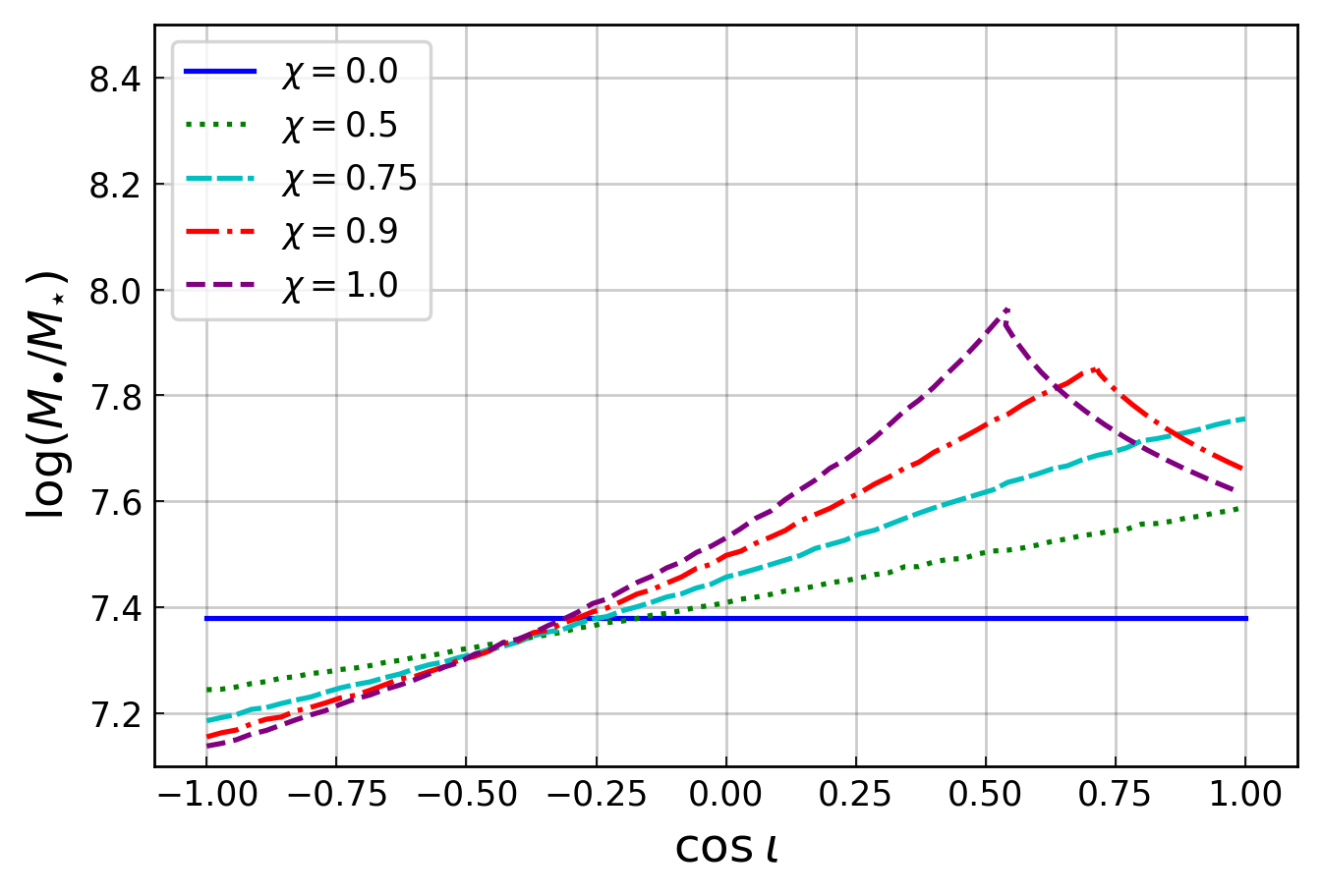}
\end{subfigure}
\caption{
The maximum SMBH mass $M_\bullet$ capable of producing a TDE for a solar-type star ($M_\star = M_\odot, R_\star = R_\odot$) as a function of inclination $\iota$.  The left panel applies in both the full loss cone (FLC) and inclination-preserving (IP) limits, while the right panel applies to the isotropizing (ISO) limit for stars with specific binding energy $\mathcal{E}^\ast = 1$.  The solid blue, dotted green, long-dashed cyan, dot-dashed red, and short-dashed purple curves correspond to SMBH spins $\chi = 0, 0.5, 0.75, 0.9$, and 1.
}
\label{fig: Max Mass}
\end{figure*}

In Fig.~\ref{fig: Max Mass}, we show the maximum SMBH mass $M_\bullet$ capable of tidally disrupting a solar-type star as a function of inclination $\iota$.  This maximum mass satisfies the equation
\begin{equation} \label{E:MaxCond}
L_d(\chi, \iota) = {\rm max}\{L_{\rm cap}(\chi, \iota), L_0(\chi, \iota)\}~.
\end{equation}
For the full loss cone (FLC), $L_0 = 0$ and this equation reduces to $L_d(\chi, \iota) = L_{\rm cap}(\chi, \iota)$.  For the steady-state loss cone in the inclination-preserving (IP) limit, $L_0(\chi, \iota) < L_{\rm lc}(\chi, \iota)$, with $L_{\rm lc}$ given by Eq.~(\ref{E:LCB}), and Eq.~(\ref{E:MaxCond}) again reduces to $L_d(\chi, \iota) = L_{\rm cap}(\chi, \iota)$. The maximum mass is thus the same in the FLC and IP limits, and is shown for both cases in the left panel of Fig.~\ref{fig: Max Mass}.  Because direct capture has a stronger retrograde bias than tidal disruption, the fully retrograde orbit ($\cos\iota = -1$) satisfies Eq.~(\ref{E:MaxCond}) at the lowest SMBH mass $M_\bullet \approx 10^{7.15} M_\odot$ for $\chi = 1$.  This can also be seen in Fig.~\ref{fig: C&D Angular momenta}, where the disruption curve $L_d(\chi = 1, \iota = \pi)$ is barely above the capture curve $L_{\rm cap}(\chi = 1, \iota = \pi)$ in the top right panel ($M_\bullet = 10^7 M_\odot$) and has fallen below it in the bottom left panel ($M_\bullet = 10^{7.25} M_\odot$).  For SMBH spins $\chi \leq 0.9$, stars on orbits with lower inclination can avoid direct capture up to higher SMBH masses, with the maximum mass for each spin occurring on the fully prograde orbit ($\cos\iota = +1$).  For the highest spin ($\chi = 1$), the capture curve $L_{\rm cap}(\chi = 1, \iota)$ has positive curvature for $\cos\iota \gtrsim 0.75$ as can be discerned with very careful scrutiny of the upper left corners of the four panels in Fig.~\ref{fig: C&D Angular momenta}.  This implies that for $M_\bullet \gtrsim 10^{8.3} M_\odot$, Eq.~(\ref{E:MaxCond}) will be satisfied on the fully prograde orbit while capture can still be avoided for other orbits with $\cos\iota \gtrsim 0.75$.  For SMBH masses $M_\bullet \gtrsim 10^{8.45} M_\odot$, tidal disruption is no longer possible, with the last surviving inclination occurring at $\cos\iota \approx 0.82$.
It is interesting to note that this corresponds to the critical inclination $\cos\iota = \sqrt{2/3}$ below which the geodesics at the capture threshold $L_{\rm cap}$ of near extremal ($\chi \approx 1$) Kerr BHs undergo a phase transition to having a finite proper radial distance to the BH horizon at pericenter \cite{Hod2013}.  Our results suggest that this phase transition may also correspond to weaker tidal forces on these orbits.

The maximum SMBH mass $M_\bullet$ capable of tidally disruption has a different dependence on inclination in the isotropizing (ISO) limit shown in the right panel of Fig.~\ref{fig: Max Mass}.  In this limit, the loss-cone boundaries $\bar{L}_{\rm lc}^2(\chi)$ given by Eq.~(\ref{E:LCB_ISO}) are independent of inclination, as are the lowest occupied orbits $L_{\rm 0,ISO}(\chi)$ given by Eq.~(\ref{E:R_0}).  They thus appear as vertical lines in the bottom panel of Fig.~\ref{fig: C&D with LC for 10^7}.  This implies that for $M_\bullet \gtrsim 10^{7.6} M_\odot$ (for specific binding energy $\mathcal{E}^\ast = 1$), $L_{\rm 0,ISO}(\chi)$ can exceed $L_{\rm cap}(\chi, \iota)$ on the right-hand side of Eq.~(\ref{E:MaxCond}) for high enough SMBH spins $\chi$ and low enough inclinations $\iota$.  This never occurs for $\cos\iota \lesssim 0.55$ implying that the curves in the left and right panels of Fig.~\ref{fig: Max Mass} agree for these values.  However, it does occur for SMBH spins $\chi = 0.9$ and 1, reducing the maximum mass capable of tidal disruption for these spins at low inclinations.  The physical interpretation of this result is that although orbits exist between the disruption and capture loss cones for these spins and inclinations, isotropization scatters stars onto higher inclination capture orbits before they can diffuse down in angular-momentum space to reach the surviving disruption orbits.  We restrict ourselves to $\mathcal{E}^\ast \geq 1$ in this paper because such binding energies simplify the loss-cone analysis and typically dominate the total event rate, but this restriction clearly breaks down in this extreme portion of parameter space.

\begin{table*}[t!]
    \centering
    \begin{tabular}{|c|c|c|c|c|c|c|c|c|c|c|c|}
    \hline
    $M_\bullet/M_\odot$ & $\chi$ & $\langle\cos\iota\rangle_{\rm FLC}$ & $\langle\cos \iota\rangle_{\rm IP}$ & $\langle\cos\iota\rangle_{\rm ISO}$ & $\log\dot{N}_{\rm IP}$ & $\log\dot{N}_{\rm ISO}$ & $\beta_{\rm min}$ & $\beta_{\rm max}$ & $\langle\beta\rangle_{\rm FLC}$ & $\langle\beta\rangle_{\rm IP}$ & $\langle\beta\rangle_{\rm ISO}$\\
    \hline
    $10^{6.5}$ & 0 & 0 & 0 & 0 & -3.7264 & -3.7264 & 1.4706 & 2.7283 & 1.9716 & 1.8029 & 1.8029 \\
    $10^{6.5}$ & 0.5 & 0.0641 & 0.0426 & -0.0109 & -3.7238 & -3.7165 & 1.4016 & 3.7451 & 1.9667 & 1.7369 & 1.7330 \\
    $10^{6.5}$ & 0.75 & 0.0984 & 0.0645 & -0.0150 & -3.7203 & -3.7044 & 1.3676 & 4.8494 & 2.0084 & 1.7172 & 1.7085 \\
    $10^{6.5}$ & 0.9 & 0.1221 & 0.0787 & -0.0160 & -3.7170 & -3.6948 & 1.3474 & 6.2984 & 2.0661 & 1.7123 & 1.6996 \\
    $10^{6.5}$ & 1.0 & 0.1474 & 0.0915 & -0.0133 & -3.7131 & -3.6868 & 1.3340 & 10.9067 & 2.1802 & 1.7226 & 1.7053\\
    \hline
    $10^7$ & 0 & 0 & 0 & 0 & -4.0956 & -4.0956 & 1.0943 & 1.2664 & 1.1762 & 1.1583 & 1.1583 \\
    $10^7$ & 0.5 & 0.2058 & 0.1527 & -0.0960 & -4.0847 & -3.9766 & 0.9641 & 1.7383 & 1.1543 & 1.0816 & 1.0641 \\
    $10^7$ & 0.75 & 0.2933 & 0.2178 & -0.0946 & -4.0709 & -3.9420 & 0.9062 & 2.2509 & 1.1746 & 1.0510 & 1.0212 \\
    $10^7$ & 0.9 & 0.3432 & 0.2528 & -0.0841 & -4.0592 & -3.9319 & 0.8736 & 2.9235 & 1.2080 & 1.0341 & 0.9969 \\
    $10^7$ & 1.0 & 0.3930 & 0.2789 & -0.0717 & -4.0474 & -3.9269 & 0.8528 & 5.0624 & 1.2865 & 1.0251 & 0.9821 \\
    \hline
    $10^{7.5}$ & 0.5 & 0.8692 & 0.8687 & 0.8518 & -5.9770 & -7.1969 & 0.4751 & 0.8069 & 0.5583 & 0.5438 & 0.5185 \\
    $10^{7.5}$ & 0.75 & 0.7703 & 0.7555 & 0.6723 & -5.0111 & -6.3589 & 0.4358 & 1.0448 & 0.5597 & 0.5315 & 0.4901 \\
    $10^{7.5}$ & 0.9 & 0.7341 & 0.7029 & 0.5474 & -4.7463 & -6.0816 & 0.4157 & 1.3570 & 0.5722 & 0.5261 & 0.4753 \\
    $10^{7.5}$ & 1.0 & 0.7375 & 0.6748 & 0.4598 & -4.6027 & -5.9160 & 0.4034 & 2.3498 & 0.6067 & 0.5230 & 0.4662 \\
    \hline
    $10^8$ & 1.0 & 0.8783 & 0.8603 & - & -5.2072 & - & 0.1872 & 1.0907 & 0.2818 & 0.2588 & - \\
    \hline
    \end{tabular}
    \caption{The third, fourth, and fifth columns list the mean cosines of the inclination $\langle\cos\iota\rangle$ for full loss cones (FLC) and steady-state loss cones refilled by inclination-preserving (IP) or isotropizing (ISO) stellar scattering for the SMBH masses $M_\bullet$ and spins $\chi$ listed in the first and second columns.  The sixth and seventh columns list the total TDE rates per galaxy per year in the IP and ISO limits.
    The eighth and nineth columns list the minimum and maximum values of the penetration factor $\beta \equiv L_t^2/L^2$ that lead to observable TDEs.  The tenth, eleventh, and twelfth columns list the mean penetration factors $\langle\beta\rangle$ in the FLC, IP, and ISO approximations.}
    \label{T: mean_u}
\end{table*}

In Table~\ref{T: mean_u}, we show the mean cosines of the inclination $<\cos\iota>$  for SMBH masses $10^{6.5} \leq M_\bullet \leq 10^8 M_\odot$ and spins $0 \leq \chi \leq 1$ for the full loss cone (FLC) and the limits of inclination-preserving (IP) and isotropizing (ISO) reflling of the loss cone.  The first ten rows of this table correspond to the SMBH masses shown in Fig.~\ref{fig: PDFs for Loss Cone, Mass}.  The IP and ISO limits shown prograde and retrograde biases respectively as explained in the discussion of that figure.  We also show the FLC limit for comparison; full exposure to the strong retrograde bias of direct capture imposes an even stronger prograde bias on the surviving disruption orbits in this case.  At $M_\bullet = 10^{7.5} M_\odot$, the TDE rate vanishes for $\chi = 0$ \cite{ServinKesden2017} and thus the mean inclinations are not shown.  For the higher spins, the FLC and IP limits have even stronger prograde biases than for lower masses and direct capture has restored the prograde bias in the ISO limit.  At $M_\bullet = 10^8 M_\odot$, tidal disruption is no longer possible for the ISO limit or $\chi \leq 0.9$ in the FLC and IP limits.  Maximally spinning SMBHs have an even stronger prograde bias than for $M_\bullet = 10^{7.5} M_\odot$.  We do not extend the table beyond $M_\bullet = 10^8 M_\odot$ because of the low TDE rates at these SMBH masses and thus do not see the reduced prograde bias that should occur for $M_\bullet \gtrsim 10^{8.3} M_\odot$ according to the left panel of Fig.~\ref{fig: Max Mass}.

\section{Penetration factor distribution}
\label{s:beta distribution}

The orbital angular momentum $L$ of the tidally disrupted star is often parameterized in terms of the penetration factor $\beta \equiv L_{t}^2/L^2$, where $L_t$ is the Newtonian threshold for tidal disruption given by Eq.~(\ref{E:L_lc N}) with structure factor $\beta_d = 1$.  The probability distribution function for the penetration factor $P(\beta)$ can be found by integrating the differential TDE rate of Eq.~(\ref{E:tripdif}) over orbital binding energies and cosine inclinations over the range $(\cos\iota)_{\rm min} \leq \cos\iota \leq (\cos\iota)_{\rm max}$, where
\begin{align}
(\cos\iota)_{\rm min} &= {\rm max}\{ -1, \cos\iota_0(\chi, L), \cos\iota_{\rm cap}(\chi, L) \}~,
\label{E:cosinc_min} \\
(\cos\iota)_{\rm max} &= {\rm min}\{ \cos\iota_d(\chi, L), +1 \}~,
\label{E:cosinc_max}
\end{align}
and $\cos\iota_0(\chi, L)$, $\cos\iota_{\rm cap}(\chi, L)$,  $\cos\iota_d(\chi, L)$ are found by inverting the lowest occupied orbit $L_0(\chi, \iota)$ and the boundaries of the loss cones $L_{\rm cap}(\chi, \iota)$ and $L_d(\chi, \iota)$ for direct capture and tidal disruption.  After multiplying by the appropriate Jacobian determinant of the coordinate transformation from $L$ to $\beta$, we obtain
\begin{equation} \label{E:P_beta}
P(\beta) \equiv \frac{d(\ln\dot{N})}{d\beta} = \frac{2r_tc^2}{GM_\bullet\beta^2} \frac{d(\ln\dot{N})}{d(L^2/M_\bullet^2)}~.
\end{equation}

\begin{figure*}[t!]
\begin{subfigure}[b]{0.49\textwidth}
\includegraphics[width=\linewidth]{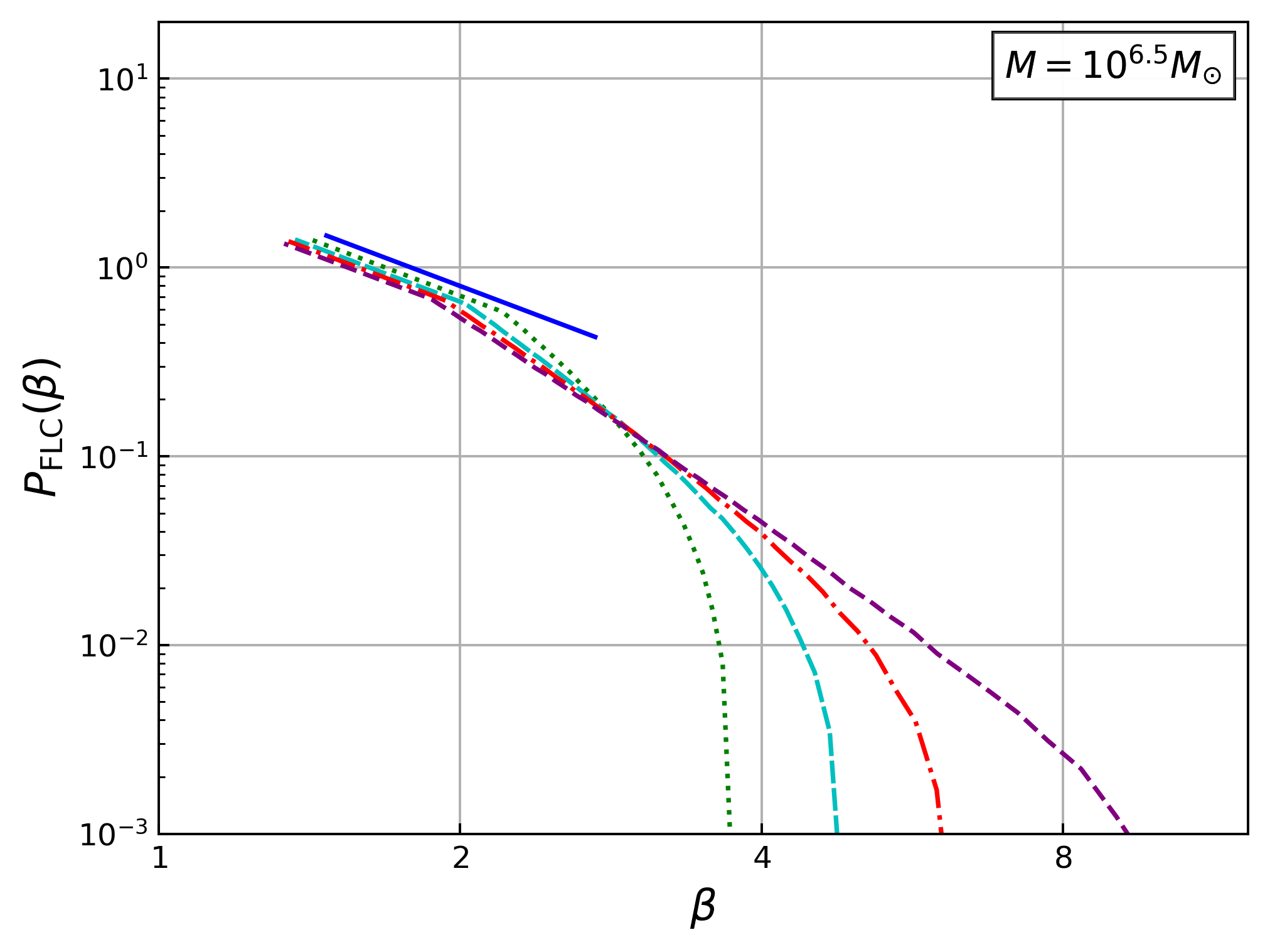}
\end{subfigure}
\begin{subfigure}[b]{0.49\textwidth}
\includegraphics[width=\linewidth]{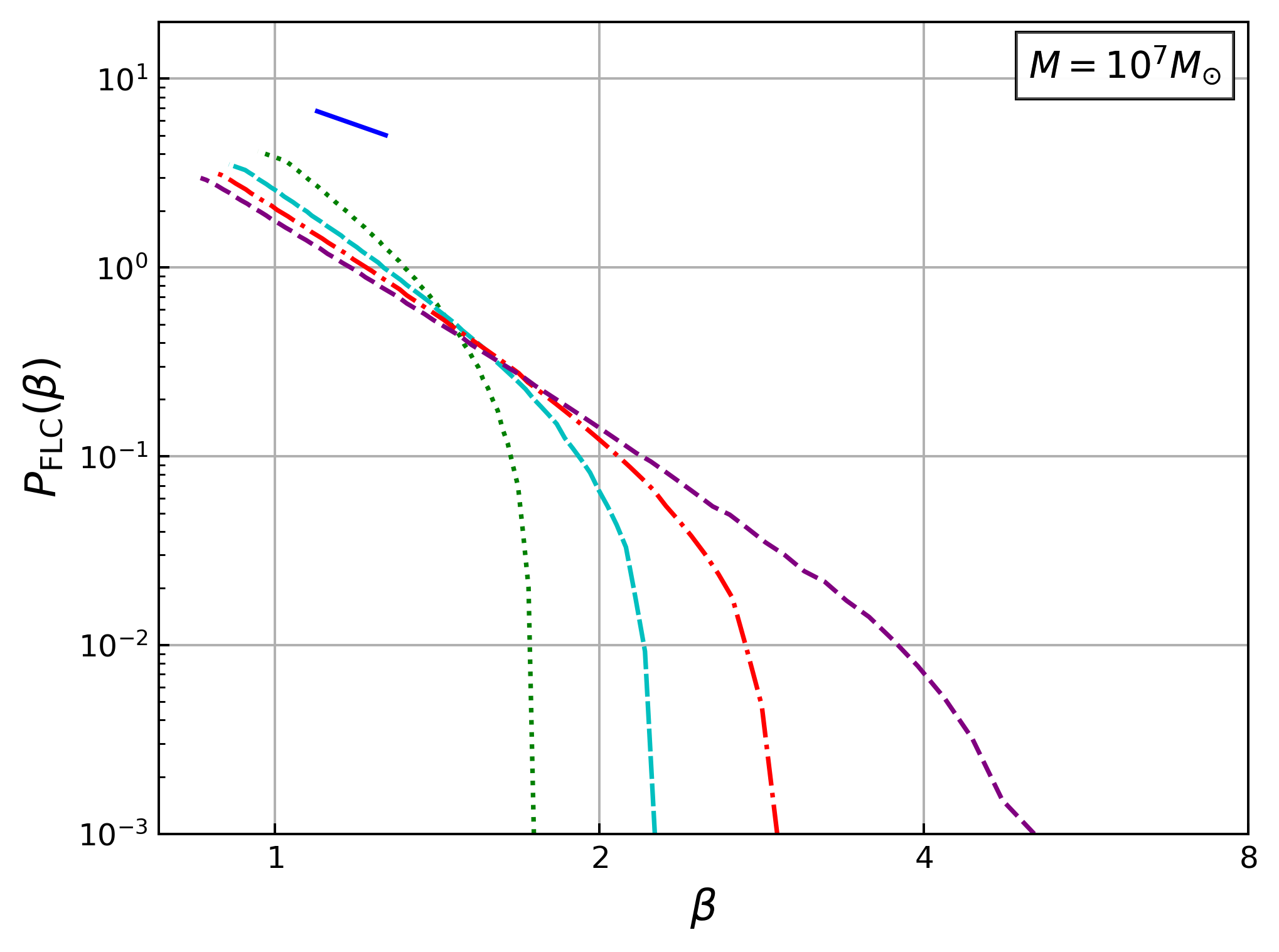}
\end{subfigure}
\begin{subfigure}[b]{0.49\textwidth}
\includegraphics[width=\linewidth]{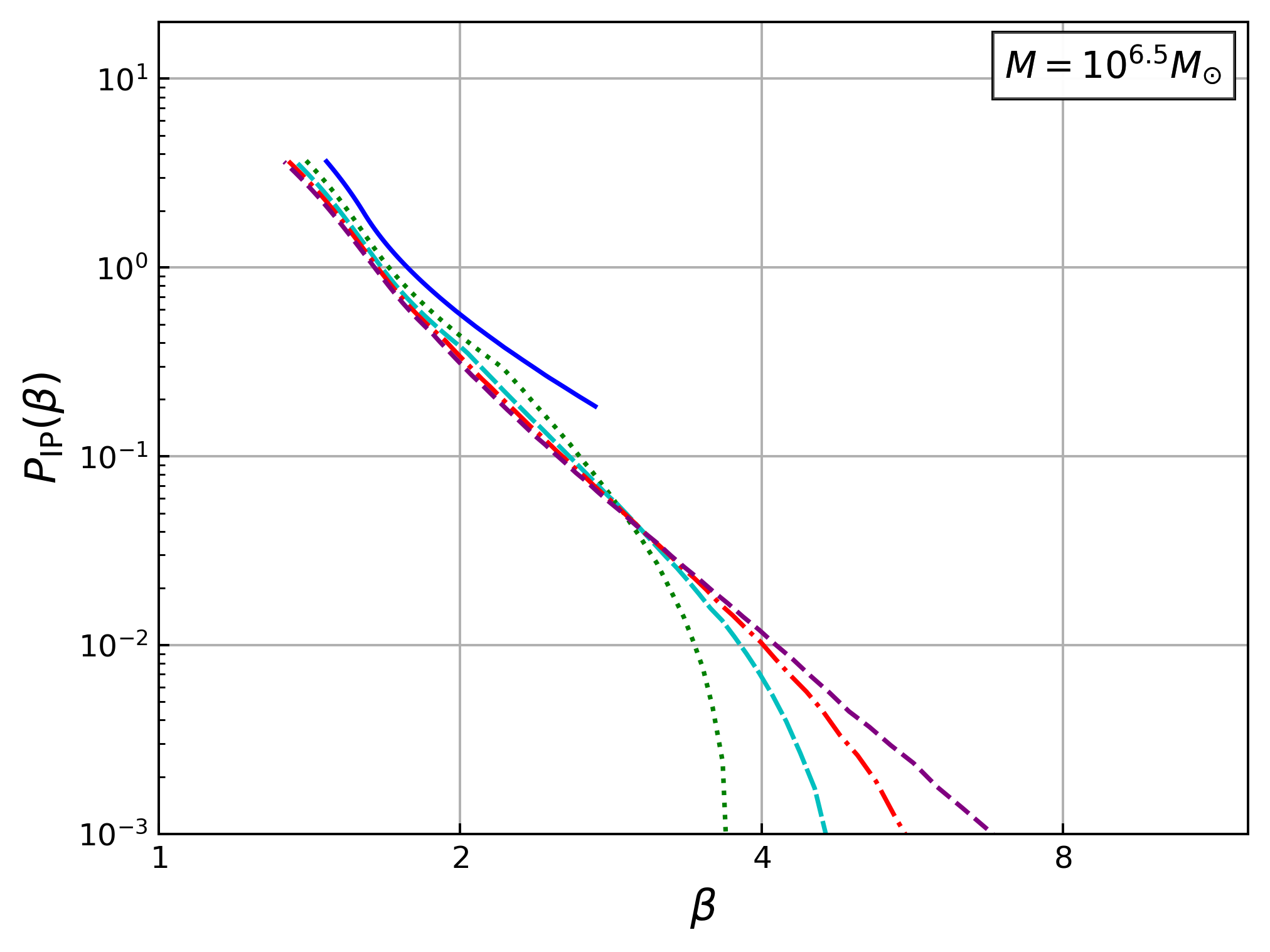}
\end{subfigure}
\begin{subfigure}[b]{0.49\textwidth}
\includegraphics[width=\linewidth]{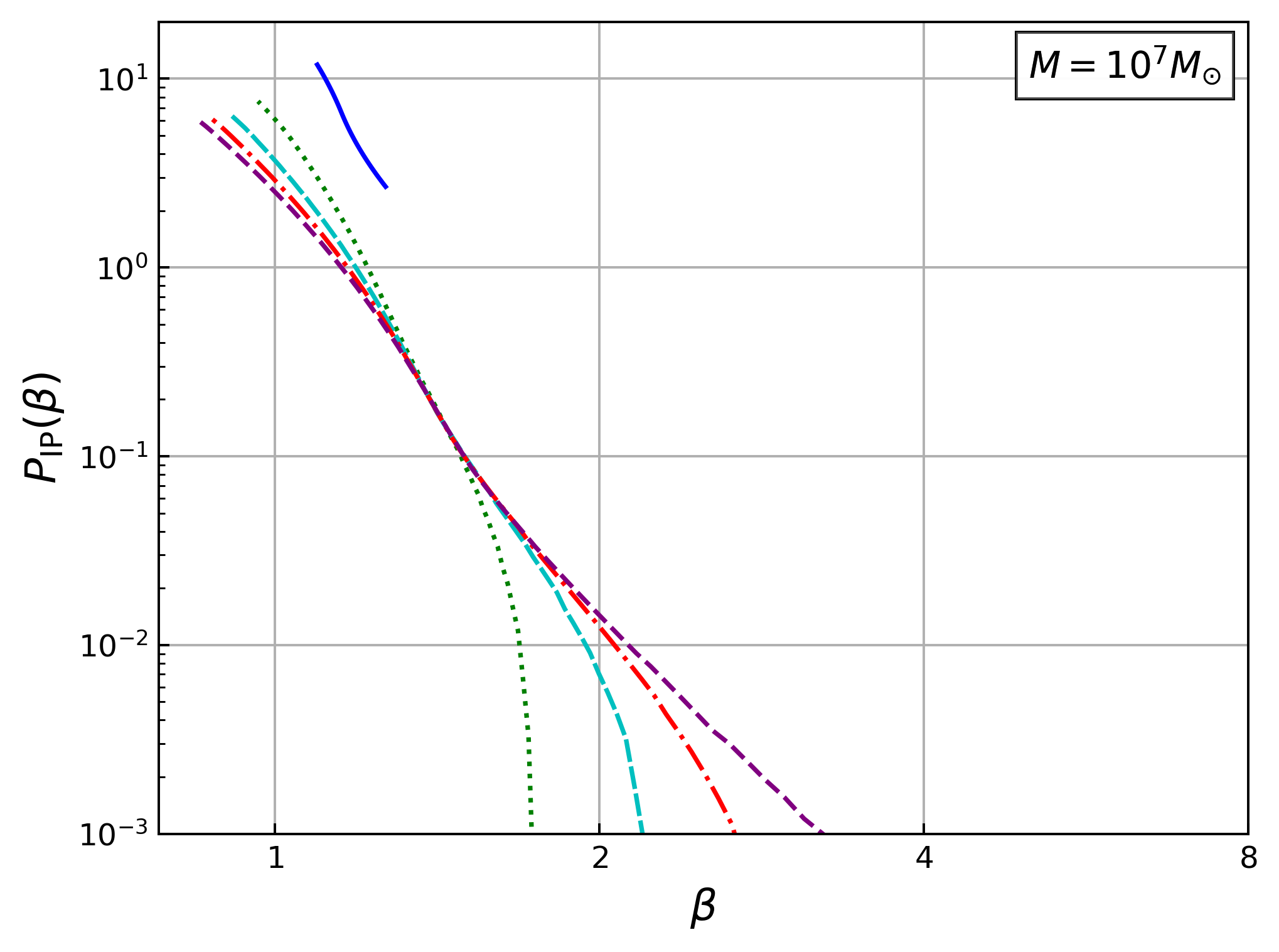}
\end{subfigure}
\begin{subfigure}[b]{0.49\textwidth}
\includegraphics[width=\linewidth]{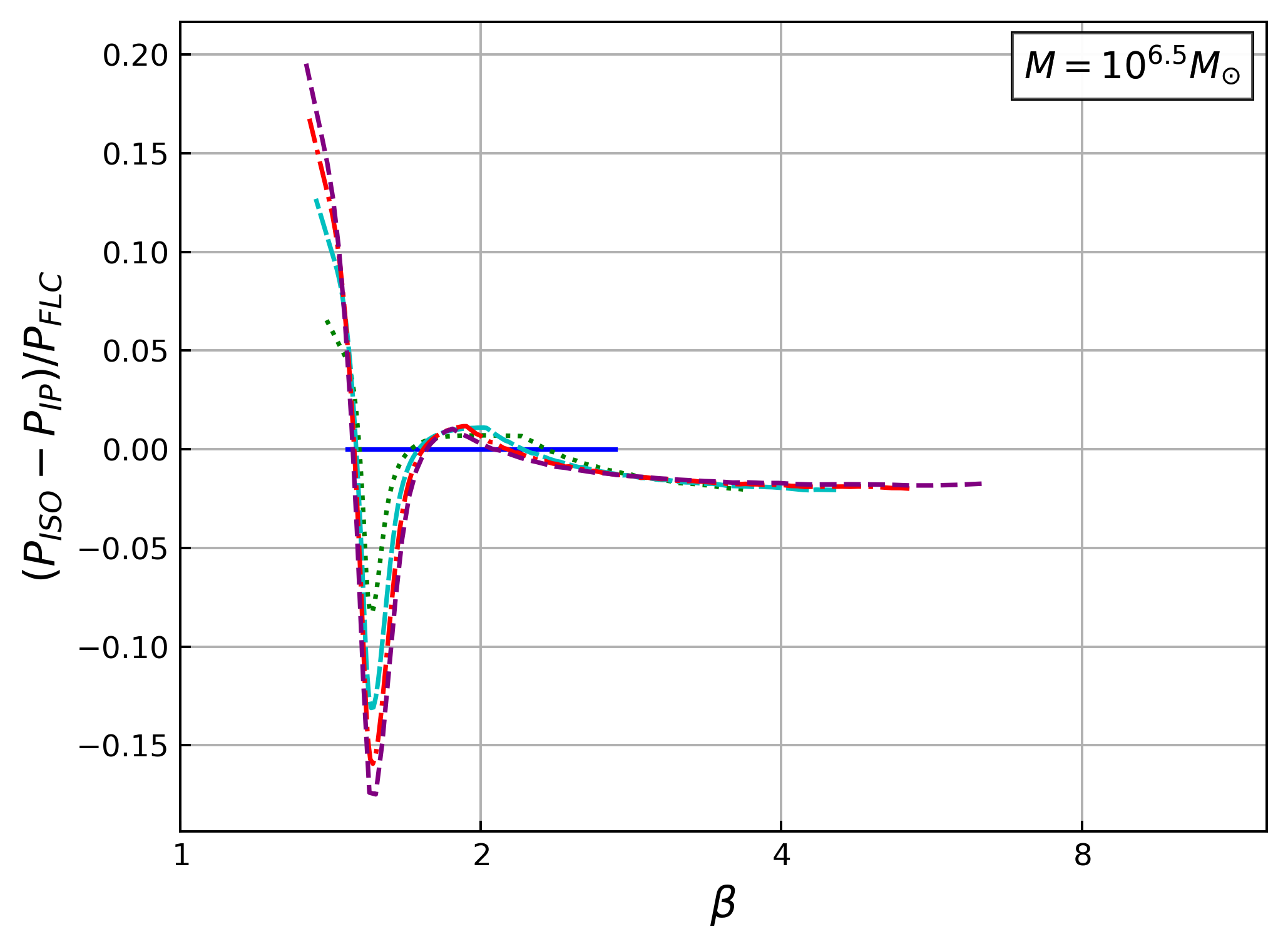}
\end{subfigure}
\begin{subfigure}[b]{0.49\textwidth}
\includegraphics[width=\linewidth]{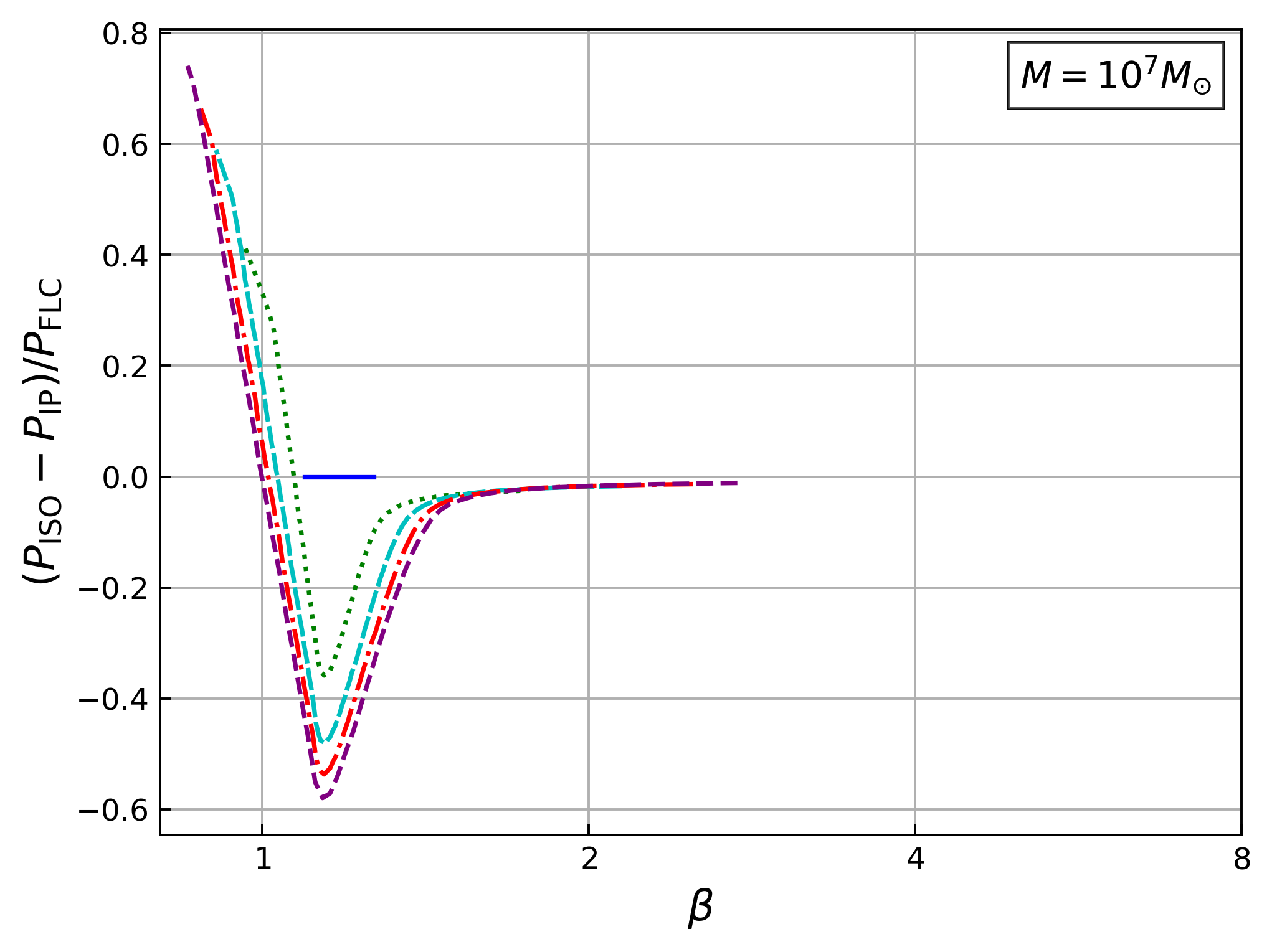}
\end{subfigure}
\caption{
The probability distribution function $P(\beta)$ of the TDE penetration factor given by Eq.~(\ref{E:P_beta}).  The left (right) panels SMBHs of mass are for SMBH mass $M_\bullet = 10^{6.5} M_\odot$ ($10^7 M_\odot$), and the solid blue, dotted green, long-dashed cyan, dot-dashed red, and short-dashed purple curves correspond to SMBH spins $\chi = 0, 0.5, 0.75, 0.9,$ and 1.  The top and middle panels show $P(\beta)$ assuming a full loss cone (FLC) and inclination-preserving (IP) refilling given by Eq.~(\ref{E:f_IP}).
In the bottom panels, we show the difference in $P(\beta)$ in the isotropizing (ISO) and IP limits normalized by the FLC prediction shown in the top panels.}

\label{fig: PDFs for beta distribution}
\end{figure*}

\subsection{Full Loss Cone}

In the FLC limit, the differential TDE rate of Eq.~(\ref{E:tripdif}) is independent of the orbital angular momentum $L$.  This implies that the penetration distribution $P(\beta)$ is determined entirely by the limits of integration given by Eqs.~(\ref{E:cosinc_min}) and (\ref{E:cosinc_max}).  We show this distribution for two different SMBH masses and a range of SMBH spins in the top panels of Fig.~\ref{fig: PDFs for beta distribution}.  In the Schwarzschild ($\chi = 0$) limit, the loss-cone boundaries are independent of inclination and $P(\beta)$ inherits the $\beta^{-2}$ dependence of the Jacobian determinant on the right-hand side of Eq.~(\ref{E:P_beta}).  This can be seen by the straight solid blue lines with a slope of -2 in the log-log plots in the top panels of Fig.~\ref{fig: PDFs for beta distribution}.  The minimum and maximum values $\beta_{\rm min}$ and $\beta_{\rm max}$ are determined by the largest and smallest values of $L$ for which the range of inclinations given by Eqs.~(\ref{E:cosinc_min}) and (\ref{E:cosinc_max}) is nonzero.  For the modest SMBH masses $M_\bullet \leq 10^7 M_\odot$ shown in Fig.~\ref{fig: PDFs for beta distribution}, 
$\beta_{\rm min} = L_t^2/L_d^2(\chi, \cos\iota = -1)$ and $\beta_{\rm max} = L_t^2/L_{\rm cap}^2(\chi, \cos\iota = +1)$ as in Fig.~\ref{fig: C&D with LC for 10^7}.  This implies that $\beta_{\rm min}$ will be a mildly decreasing function of SMBH mass $M_\bullet$ because relativistic tides are stronger than their Newtonian counterparts \cite{ServinKesden2017}, but $\beta_{\rm max} \propto M_\bullet^{-2/3}$.  As the SMBH spin $\chi$ increases, $\beta_{\rm min}$ slightly decreases because of the mildly stronger tides on retrograde orbits, but $\beta_{\rm max}$ will greatly increase because of the strong prograde bias against direct capture at high spins.  The top panels of Fig.~\ref{fig: PDFs for beta distribution} are similar to penetration factor distribution shown in the right panel of Fig.~3 of \citeauthor{CoughlinNixon2022}~\cite{CoughlinNixon2022}, although our coordinate-independent definition of $\beta$ differs from their definition $\beta_{\rm CN} \equiv r_t/r_p$ in terms of the Boyer-Lindquist radial coordinate at pericenter $r_p$.  We list $\beta_{\rm min}$, $\beta_{\rm max}$, and the mean penetration factor $\langle\beta\rangle$ under different assumptions about the occupation of the loss cone in Table.~\ref{T: mean_u}.

\subsection{Steady-State Loss Cone}

The middle panels of Fig.~\ref{fig: PDFs for beta distribution} show the penetration factor distribution $P(\beta)$ assuming the inclination-preserving (IP) steady-state phase-space distribution function of Eq.~(\ref{E:f_IP}).  Although $\beta_{\rm min}$ and $\beta_{\rm max}$, set by the loss-cone boundaries, have not changed, $P(\beta)$ has steepened considerably compared to the FLC limit shown in the top panels.  Because stars have less than an orbital period to diffuse within the loss cone before being disrupted at $r < r_t$, it is more difficult for them to reach larger values of $\beta$.  This statement is equivalent to saying that as $q \to 0$ in Eq.~(\ref{E:q}), $\mathcal{R} \to \mathcal{R}_0$ in Eq.~(\ref{E:R_0}) and thus $L \to L_d$, $\beta \to L_t^2/L_d^2 \to \beta_{\rm min}$.  The reduction in the mean penetration factor $\langle\beta\rangle$ in the IP limit compared to the FLC limit as a function of SMBH mass and spin can also be seen in Table.~\ref{T: mean_u}.

Although the assumption of inclination-preserving (IP) versus isotropizing (ISO) refilling of the loss cone strongly affects the inclination distributions as shown in Fig.~\ref{fig: PDFs for Loss Cone, Mass}, by design Eq.~(\ref{E:LCB_ISO}) preserves the inclination-averaged of $\mathcal{R}_0$ given by Eq.~(\ref{E:R_0}) and thus the inclination-averaged rate at which orbits within the loss cone are refilled.  This implies that the penetration factor distribution $P(\beta)$ will be largely insensitive to this assumption.  In the bottom panels of Fig.~\ref{fig: PDFs for beta distribution}, we show the difference in $P(\beta)$ under these two assumptions normalized by the FLC result.  We see that the ISO limit leads to an even steeper penetration factor distribution than the IP limit near $\beta_{\rm min}$.  At the smallest values of $\beta$, disruption is only occurring on retrograde orbits because of the stronger tidal forces on such orbits as indicated by the $L_d(\chi, \iota)$ curves shown in Fig.~\ref{fig: C&D Angular momenta}.  The excess of disruptions and these values of $\beta$ seen in the bottom panels of Fig.~\ref{fig: PDFs for beta distribution} is thus simply another representation of the retrograde bias in the ISO limit seen in the right panels of Fig.~\ref{fig: PDFs for Loss Cone, Mass}.  As $\beta$ increases, tides become strong enough to disrupt stars on prograde orbits as well and $P(\beta)$ includes contributions from these orbits on which the phase-space distribution function is comparatively suppressed in the ISO limit according to Eq.~(\ref{E:f_ISO}).  As $\beta$ is increased further and $L$ approaches the lowest occupied orbit $L_0$, the loss cone is emptied in both the ISO and IP limits and this the ratio shown in the bottom panels approaches zero.

\section{Total Rates} \label{S:Total Rates}

\begin{figure*}[t!]
\begin{subfigure}[b]{0.49\textwidth}
\includegraphics[width=\linewidth]{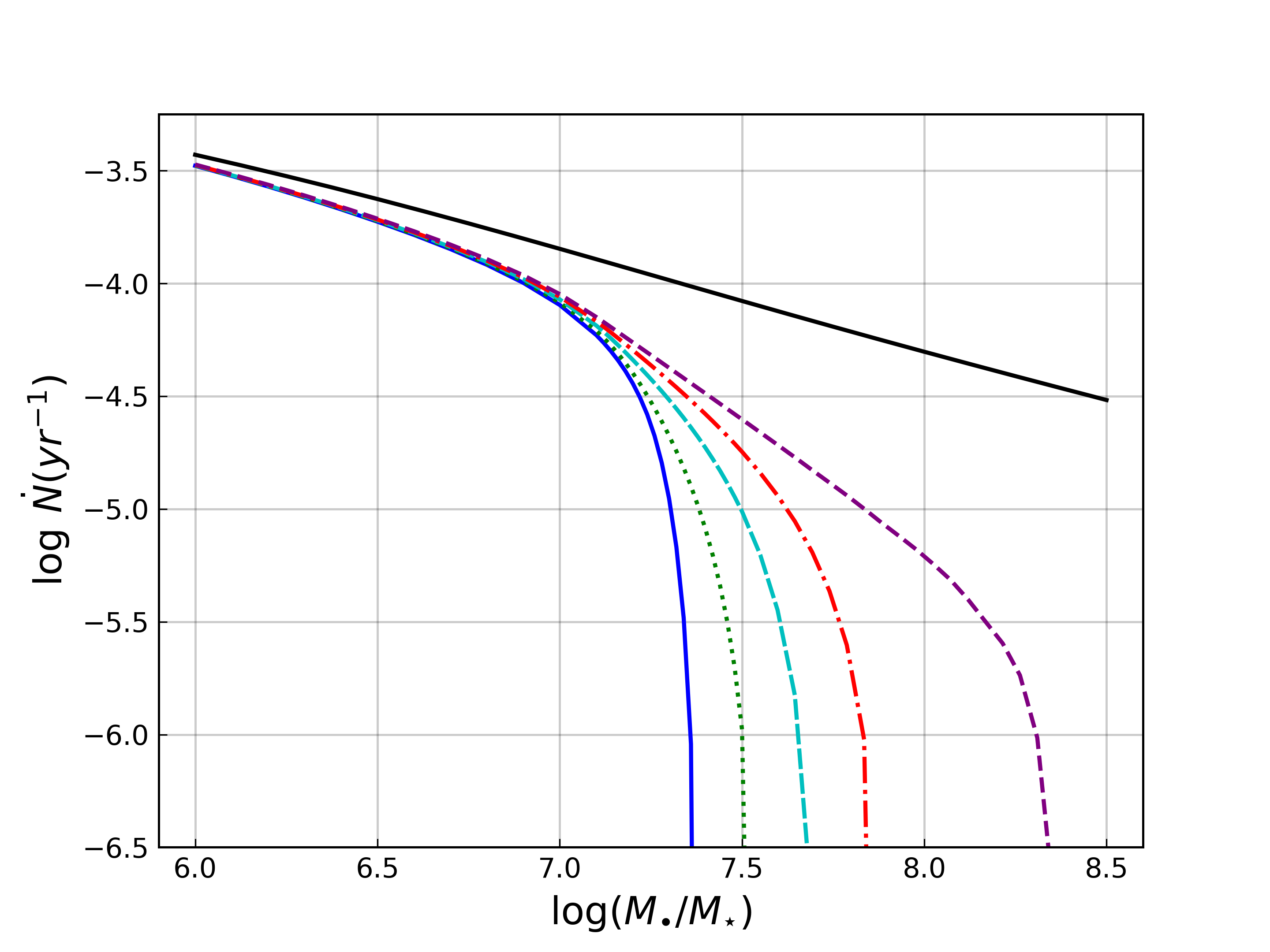}
\end{subfigure}
\begin{subfigure}[b]{0.49\textwidth}
\includegraphics[width=\linewidth]{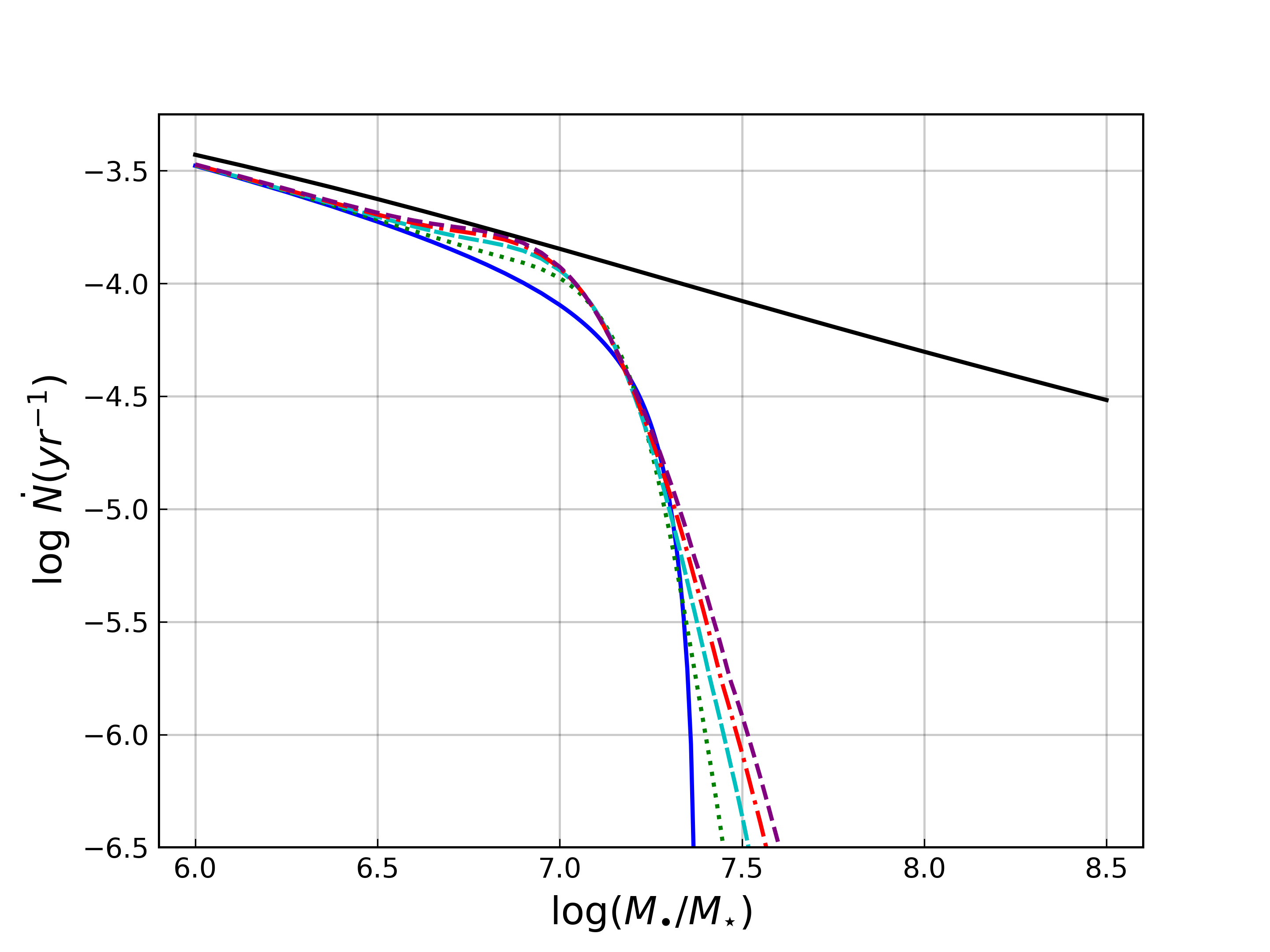}
\end{subfigure}
\caption{Total TDE rate per galaxy $\dot{N}$ as a function of SMBH mass $M_\bullet$.  The left (right) panels show the inclination-preserving (IP) and isotropizing (ISO) limits of the steady-state loss cone.  The solid blue, dotted green, long-dashed cyan, dot-dashed red, and short-dashed purple curves correspond to spins $\chi = 0, 0.5, 0.75, 0.9$, and 1, while the solid black curve shows the Newtonian prediction.
}
\label{fig: TDE Rates}
\end{figure*}

Although this paper has focused on the inclination distribution function $P(\cos\iota)$, for completeness, we also calculate the total TDE rate
\begin{align}
\dot{N}(M_\bullet, \chi) &= \int_1^\infty d\mathcal{E}^\ast \int_{-1}^{+1} d(\cos\iota) \notag \\
&\qquad \times \int_{L_{\rm cap}^2/M_\bullet^2}^{L_d^2/M_\bullet^2} d\left( \frac{L^2}{M_\bullet^2} \right) \frac{d^3\dot{N}}{d\mathcal{E}^\ast d(L^2/M_\bullet^2)d(\cos\iota)}~, \label{E:totrates}
\end{align}

where the integrand vanishes for $L_d < {\rm max}\{L_{\rm cap}, L_0 \}$.  This total rate is shown as a function of SMBH mass $M_\bullet$ in both the IP and ISO limits in Fig.~\ref{fig: TDE Rates} and for selected SMBH masses and spins in Table~\ref{T: mean_u}.  For comparison, we also show the Newtonian prediction for the total TDE rate in  Fig.~\ref{fig: TDE Rates} which includes neither of the relativistic effects of geodesic deviation or direct capture by the event horizon.

For $M_\bullet \lesssim 10^{6.5} M_\odot$, we see a mild suppression in the total TDE rate due to direct capture in both the IP and ISO limits, despite the fact that relativity can already impose significant inclination biases at these masses, particularly in the IP limit, as shown by the top panels of Fig.~\ref{fig: PDFs for Loss Cone, Mass}.  The symmetry in these inclination distributions about $\cos\iota = 0$ implies that the spin dependence of the integrated effect on the total rate of Eq.~(\ref{E:totrates}) is negligible.

For $10^{6.5} M_\odot \lesssim M_\bullet \lesssim 10^7 M_\odot$, we see a further reduction in the total TDE rate compared to the Newtonian prediction in the IP limit.  Its weak dependence on SMBH spin is consistent with prior predictions for the total capture rate \cite{Young1977}.  However, for spins $\chi \geq 0.5$ in the ISO limit, we see an \emph{enhancement} in the TDE rate above the Schwarzschild values that almost reaches the horizon-free Newtonian prediction.  This enhancement is a new result that to our knowledge has not been reported in any prior work.   It occurs because the occupation of the loss cone increases with $L$ according to Eq.~(\ref{E:f_elc}), implying that the gain in retrograde TDEs due to isotropization of the loss cone exceeds the loss in prograde TDEs in this mass range.  The enhancement has a mild spin dependence at peaks at $\sim 75\%$ over the Schwarzschild prediction at $M_\bullet \approx 10^{6.9} M_\odot$.

For $M_\bullet \gtrsim 10^7 M_\odot$, direct capture dominates, and the total TDE rate is sharply reduced below the Newtonian prediction.  In the IP limit, the cutoff spans over an order of magnitude, from $10^{7.39} M_\odot$ for $\chi = 0$ all the way up to $10^{8.45} M_\odot$ for $\chi = 1$, as can be seen from the left panels of Figs.~\ref{fig: Max Mass} and \ref{fig: TDE Rates}.  This broad range occurs because of the strong spin dependence of the retrograde bias of direct capture which allows stars on prograde orbits to survive capture up to high SMBH masses.  In the ISO limit, relativity giveth and relativity taketh away.  The cutoff occurs at lower masses and is much less spin dependent because the isotropization of the loss cone suppresses precisely those prograde orbits essential to TDEs at high SMBH masses.

The left panel of Fig.~\ref{fig: TDE Rates} qualitatively resembles Fig.~4 of Kesden~\cite{Kesden2012}, but there are several differences between that calculation and the one presented here.  Comparing Eq.~(11) of that paper to Eq.~(\ref{E:DisThresh}) of this one reveals that the earlier paper implicitly assumed a Newtonian threshold for tidal disruption of $\beta_d = 2^{-1/3} \approx 0.79$ much smaller than the value $\beta_d \approx 1.9$ found by Guillochon and Ramirez-Ruiz~\cite{Guillochon2013} to be appropriate for the full disruption of a solar-type star.  The smaller value of $\beta_d$ in the earlier paper implied that the stars were more susceptible to tidal disruption (at a smaller penetration factor), shifting the cutoffs to lower masses.  Kesden~\cite{Kesden2012} also did not self-consistently include loss-cone physics as in this paper.  Instead, it calculated TDE rates assuming a full loss cone, then normalized these rates at each SMBH mass to agree with the Newtonian suppression of the rate by the partially empty loss cone in that limit.  This normalization procedure would effectively overestimate the influence of direct capture at intermediate SMBH masses where the empty loss cone would suppress capture but not disruption, i.e. $L_0 \simeq L_{\rm cap} < L_d$.  Finally, Kesden~\cite{Kesden2012} used an older version \cite{SchulzeGebhardt2011} of the $M_\bullet-\sigma$ relation between SMBH mass and host-galaxy velocity dispersion than the one \cite{McConnellMa2013} employed here.

\section{Discussion} \label{S:Discussion}

In this paper, we investigated the interplay between relativistic tidal disruption, direct capture by the event horizon, and loss-cone physics in determining the TDE inclination distribution function, penetration factor distribution function, and total event rate for spinning SMBHs.  Although TDE inclination and penetration factor are not directly observable, they are important because of their potential effects on tidal-stream circularization \cite{Kochanek1994,StoneLoeb2012,Guillochon2015,BatraLu2021}, the ability of TDEs to launch relativistic jets \cite{Burrows2011,Cenko2012,Brown2015,BlandfordZnajek1977,2023arXiv230805161T}, the the amplitude and frequency of quasi-periodic oscillations (QPOs) observed in TDE candidates \cite{Reis2012,Lin2015,Pasham2019,StoneLoeb2012,vanVelzenPasham2021}, and the total energy reservoir available at the innermost stable circular orbit (ISCO) \cite{BardeenTeukolsky1972}.  Both tidal disruption and capture have a retrograde bias, i.e. the threshold of specific orbital angular momenta $L_d(\chi, \iota)$ and $L_{\rm cap}(\chi, \iota)$ below which each process occurs is a monotonically increasing function of inclination $\iota$.  As capture has a stronger bias than disruption and avoiding capture is a necessary condition for producing an observable TDE, the prevailing wisdom is that TDEs should have a prograde bias.  

This wisdom holds in the full loss cone (FLC) limit, i.e. when the orbits on which tidal disruption or capture occur are repopulated by new stars from the galactic center of the host galaxy on timescales much shorter than the orbital time.  This FLC limit is a reasonable approximation for small SMBHs ($M_\bullet \lesssim 10^6 M_\odot$) because of their smaller loss cones and denser stellar environments, but breaks down for higher masses where relativistic effects become significant $L_{\rm cap}/L_d \propto M_\bullet^{1/3}$.  For spherically symmetric Newtonian potentials, the Fokker-Planck equation can be orbit-averaged in energy-angular momentum phase space to determine the steady-state TDE rate when the loss cone is partially empty \cite{CohnKulsrud1978,MagorrianTremaine1999,WangMerritt2004,StoneMetzger2016}.  As the stellar scattering responsible for refilling the loss cone occurs far from the SMBH where relativistic effects can be neglected, this same approach can be applied to determine TDE rates by non-spinning Schwarzschild SMBHs \cite{ServinKesden2017}.

SMBH spin breaks the spherical symmetry of the Schwarzschild spacetime, invalidating the averaging procedure used in previous TDE rate calculations based on the Fokker-Planck equation.  A new anisotropic averaging procedure could potentially be developed to remedy this situation, but we wanted to examine the potential consequences of anisotropic occupation of loss-cone orbits before undertaking such a demanding investigation.  In this work, we consider the two extreme limits: inclination preserving (IP), in which each inclination $\iota$ is approximated as uncoupled with its own loss-cone boundary $L_{\rm lc}(\chi, \iota)$ given by Eq.~(\ref{E:LCB}), and isotropizing (ISO), in which diffusion in phase space is assumed to fully isotropize the occupation of the loss cone with an inclination-averaged phase-space distribution function give by Eq.~(\ref{E:f_ISO}).

We discovered that the degree of loss-cone isotropization has a strong effect on the TDE inclination distribution function which can be understood by considering the hierarchy of $L_d$, $L_{\rm cap}$, and the lowest occupied orbit $L_0$.  At small SMBH masses $M_\bullet \lesssim 10^6 M_\odot$, $L_0 \ll L_{\rm cap} \ll L_d$ and both the IP and ISO limits resemble the FLC limit.  This yields prograde biases as in the top panels of Fig.~\ref{fig: PDFs for Loss Cone, Energy}.  As $M_\bullet$ increases, $L_d$ approaches $L_{\rm cap}$ as shown in the top panels of Fig.~\ref{fig: C&D Angular momenta}, reducing the allowed phase space for retrograde TDEs and thus tending to increase the prograde bias.  However, according to Eqs.~(\ref{E:R_0}) and (\ref{E:q}), $L_0$ increases as well with SMBH mass.  In the IP limit, this can at best flatten the distribution of $\cos\iota$ for $L_{\rm cap}(\chi, \iota) < L_{\rm 0, IP}(\chi, \iota) < L_d(\chi, \iota)$ as seen in the middle and lower left panels of Fig.~\ref{fig: PDFs for Loss Cone, Energy}.  It fails to overcome the dominant increase in prograde bias that characterizes the FLC limit.  However, in the ISO limit, the independence of $L_{\rm 0, ISO}(\chi)$ on inclination can conceal the strong retrograde bias of capture for $L_{\rm cap}(\chi, \iota) < L_{\rm 0, ISO}(\chi) < L_d(\chi, \iota)$.  The milder retrograde bias of disruption is thus revealed as seen in the middle and lower right panels of Fig.~\ref{fig: PDFs for Loss Cone, Energy}.  At SMBH masses $M_\bullet \gtrsim 10^{7.25} M_\odot$ as seen in the bottom panels of Fig.~\ref{fig: C&D Angular momenta}, $L_{\rm cap}(\chi, \iota) > L_d(\chi, \iota)$ for retrograde inclinations and the prograde bias is enhanced (restored) in the IP (ISO) limit.

We also explored how our different assumption about the efficiency with which the loss cone is refilled affect the distribution function of the penetration factor $\beta \equiv L_t^2/L^2$.  Once they enter the loss cone, stars have less than an orbital period to diffuse in angular-momentum space before they are disrupted at $r < r_t$.  This implies that when the loss cone is refilled with finite efficiency as in both the IP and ISO limits, the $\beta$ distribution will be steeper than in the FLC limit.  It will be even steeper in the ISO limit than the IP limit because isotropization resulting from diffusion in inclination $\iota$ in addition to angular momentum $L$ allows more stars to be disrupted on low $\beta$ retrograde orbits where relativistic effects enhance the strength of tides.  As these low $\beta$ orbits are more likely to lead to partial rather than full tidal disruptions \cite{Guillochon2013}, we expect a greater ratio of partial to full disruptions in the ISO and IP limits than in the FLC limit.

This SMBH mass-dependent hierarchy of $L_d$, $L_{\rm cap}$, and $L_0$ also effects the total TDE rate as seen in Fig.~\ref{fig: TDE Rates}.  In the IP limit, observable TDEs are possible as long as $L_d(\chi, \iota) > L_{\rm cap}(\chi, \iota)$ at \emph{any} inclination.  As seen in the left panel of Fig.~\ref{fig: Max Mass}, this condition is satisfied at $\cos\iota \approx 0.82$ for SMBH masses as large as $10^{8.45} M_\odot$ for $\chi = 1$.  In the ISO limit, the increase in the phase-space distribution function $f_{\rm ISO}$ of Eq.~(\ref{E:f_ISO}) with $L$ implies that the same hierarchy $L_{\rm cap}(\chi, \iota) < L_{\rm 0, ISO}(\chi) < L_d(\chi, \iota)$ that generated the retrograde inclination bias enhances the total TDE rate.  However, for SMBH masses $M_\bullet \gtrsim 10^{7.25} M_\odot$ for which $L_{\rm cap}(\chi, \iota) > L_d(\chi, \iota)$ on retrograde orbits, the coupling between inclinations given by Eq.~(\ref{E:LCB_ISO}) implies that the TDE rate is suppressed at \emph{all} inclinations.  This leads to a very sharp cutoff in the total TDE rate above $M_\bullet \approx 10^{7.5} M_\odot$ for all spins and a complete suppression of TDEs above $10^8 M_\odot$ as seen in the right panel of Fig.~\ref{fig: Max Mass}.

This study has revealed previously unanticipated richness in the dependence of TDE inclination distribution on the degree of loss-cone isotropization as a function of SMBH mass $M_\bullet$ and spin $\chi$.  In future work, we intend to determine this loss-cone isotropization by a proper calculation based on the orbit-averaged Fokker-Planck equation.  We will also explore how these TDE inclination distributions can be used to predict more directly observable TDE properties.  It is our fervent hope that these predictions will prove valuable in the interpretation of the many TDE candidates expected to be found in future surveys such as the Legacy Survey of Space and Time (LSST) \cite{LSST} and that undertaken by the Ultraviolet Transient Astronomy Satellite (ULTRASAT) \cite{ULTRASAT}.

\acknowledgements

The authors were supported by NASA award number 80NSSC18K0639.  They would also like to thank Juan Servin and Joseph Rossi, whose dissertation research at the University of Texas at Dallas (UTD) was supported by this award, and Thomas DeMastri and Alison Powell, who were supported as Research Experiences for Undergraduates (REU) Fellows at UTD by NSF awards PHY-1757503 and PHY-2050781.

\bibliography{main}

\end{document}